\documentclass[a4paper,12pt]{article}
\usepackage{amsmath, empheq}
\usepackage{amsfonts}
\usepackage{amssymb}
\usepackage{marvosym}
\usepackage{wasysym}
\usepackage[font=small,labelfont=bf]{caption}
\usepackage{subcaption}
\usepackage{fullpage}
\usepackage[]{graphicx}
\usepackage{hyperref}
\usepackage[english]{babel}
\usepackage{xcolor}
\usepackage{makeidx}
\newcommand{\be}{\begin{eqnarray}}
\newcommand{\ee}{\end{eqnarray}}

\def\mn{{\mu\nu}}
\def\mpl{m_{Pl}}
\newcommand{\rar}{\rightarrow}

\numberwithin{equation}{section}

\newcommand{\bi}{\bibitem}

\hypersetup{citebordercolor=0 1 0, linkbordercolor=1 1 1,}

\begin{document}

\begin{titlepage}
\title{Distortion of the standard cosmology in $R+R^2$ theory} 
\author{E.V. Arbuzova$^{1,2}$,  A. D. Dolgov$^{2,3}$, and R. S. Singh$^{2}$}
\date{}
\maketitle
\begin{center}
$^{1}$\emph{Department of Higher Mathematics,University Dubna, \\Universitetskaya ulitsa, 19, 141980 Dubna, Russia}\\
$^{2}$\emph{Department of Physics, Novosibirsk State University, \\Pirogova 2, Novosibirsk 630090, Russia\\
$^{3}${ITEP}, Bol. Cheremushkinskaya 25, Moscow 117218 Russia} \\

\end{center}
\thispagestyle{empty}

\abstract{Universe history in $R^2$-gravity is studied from "beginning" up to the present epoch. It is assumed that initially the 
curvature scalar $R$ was sufficiently large to induce the proper duration of inflation. Gravitational particle production 
by the oscillating $R(t)$ led to a graceful exit from inflation, but the cosmological evolution in the early universe was 
drastically different from the standard one till the universe age reached the value of the order of the inverse decay rate of the
oscillating curvature $R(t)$. This deviation from the standard cosmology might  have a noticeable impact on the formation 
of primordial black holes and baryogenesis. At later time, after exponential decay of the curvature oscillations, cosmology 
may return to normality.}

\vspace{5cm}
\Email{A.D. Dolgov:  \href{mailto:dolgov@fe.infn.it}{\nolinkurl{dolgov@fe.infn.it}} }\\

\Email{E.V. Arbuzova: \href{mailto:arbuzova@uni-dubna.ru}{\nolinkurl{arbuzova@uni-dubna.ru}} }\\

\Email{R.S. Singh: \href{mailto:akshalvat01@gmail.com}{\nolinkurl{akshalvat01@gmail.com}} }\\

\end{titlepage}

\pagenumbering{arabic}
\section{Introduction \label{s-Intro}}

Theory of gravitational interaction,  General Relativity (GR), based on the Einstein-Hilbert action~\cite{Hilbert,Einstein}
\be 
S_{EH}= - \frac{m_{Pl}^2}{16\pi} \int d^4 x \sqrt{-g}\,R,
\label{S-EH}
\ee
describes basic  properties of the universe in very good agreement with observations. Here $R$ is the curvature 
scalar, $m_{Pl}= 1. 2 2\cdot 10^{19}$ GeV is the Planck mass, which is connected with the gravitational coupling constant 
as $m_{Pl}^2 = G_N^{-1}$, $g$ is the determinant  of the metric tensor $g_{\mu\nu}$ with the signature convention 
$(+,-,-,-)$. The Riemann tensor describing the curvature of space-time is determined according to 
$R^\alpha_{\,\,\mu\beta\nu}=\partial_\beta\Gamma^\alpha_{\mu\nu}+\cdots$, 
$R_{\mu\nu} = R^\alpha_{\,\,\mu\alpha\nu}$, and
$R=g^{\mu\nu}R_{\mu\nu}$. We use here the natural system of units $\hbar=c=k_B=1$.

However, some features of the  universe may request to go beyond  the frameworks of GR. Usually it is achieved by 
an addition of a nonlinear function $F(R)$ into the action $S_{EH}$ (\ref{S-EH}):
\be
S_F =  -\frac{m_{Pl}^2}{16\pi} \int d^4 x \sqrt{-g}\,\left[R+F(R)\right]\, .
\label{S-F}
\ee
In 1979 V.Ts. Gurovich and A.A. Starobinsky~\cite{Gur-Star} suggested to take $F(R)=-R^2/(6m^2)$ for elimination of 
cosmological singularity.  In the subsequent paper by Starobinsky~\cite{Starobinsky_1980} it was found that the 
addition of the $R^2$-term leads to 
inflationary cosmology. As in any cosmological scenario the problem of graceful exit from inflation and the 
problem of the universe heating are of primary importance. They were studied in 
Refs.~\cite{Starobinsky_1980}-\cite{ADR-R2} { and reviewed in \cite{R2-rev}.}

In the present work we generalize and extend the analysis of our earlier paper~\cite{ADR-R2} { starting from the inflationary
stage and continuing to asymptotically  large time, such that $m t \gg 1$. First we have studied the case of large but not too large
time such that   $\Gamma t < 1$,
where $\Gamma$ is the width of the scalaron decay, see below, Eq.~(\ref{ddot-R-Gam}). In this time range one can find
very simple analytical expressions for $H(t)$ and $R(t)$, given by Eqs.~(\ref{hsol}) and (\ref{rsol}). 
These solutions are presented in review~\cite{R2-rev}, but they simply follow from the expression for the cosmological scale
factor earlier derived in ref.~\cite{Starobinsky_1980}. According to this work
the cosmological scale factor in $R^2$-theory evolves as
\be
a(t) \sim t^{3/2} \left[ 1 + \frac{2}{3 mt} \,\sin (mt + \theta)\right]
\label{a-of-t-R}
\ee
and the Hubble parameter presented in~\cite{Starobinsky_1982}  is the same as derived later by different methods in 
review~\cite{R2-rev} and here.

Below we reproduced the original result of ref.~\cite{Starobinsky_1980}, using another analytical method. In ref.~\cite{Starobinsky_1980}
the system of the cosmological equations  (in absence of the usual matter) was  transformed into a single first order non-linear equation,
while here the system of the second order equation for  $R$ (\ref{ddot-R}), the covariant law of conservation of the matter energy 
density (\ref{dot-rho}), and the "kinematical" relation between the curvature scalar and the  Hubble parameter (\ref{R-of-H}) in spatially
flat universe are employed. This system is easier to treat numerically and we want to keep the matter effects from the very beginning, 
though they are quite weak initially. We found numerically
that the onset of the simple asymptotic behavior (\ref{hsol}) and (\ref{rsol}) started almost immediately after inflation was over.
We have also calculated the energy density of the usual matter, which drops down as $1/t$ with some weak superimposed 
oscillation. At the time range such that $\Gamma t <1$ the usual matter has very weak impact on the cosmological expansion which
is determined by the oscillating curvature.
During this time interval the universe evolution is quite different from the General Relativity (GR) one. 

Though the curvature scalar in many respects reminds a usual scalar field and the expansion regime is rather similar to
the matter dominated one, there still a considerable difference between cosmological
 evolution in $R^2$- modified gravity and that induced by a homogeneous  massive scalar field $\phi$
 with mass $m_\phi$. The study of the latter was pioneered
 by Starobinsky in ref.~\cite{AAS-78}. As it is shown there, the energy density of the scalar drops down basically as $1/t^2$
with some oscillating terms decaying as  $1/t^2$ with respect to the dominant term, i.e the oscillating part  drops down as $1/t^4$.
The scale factor in this model behaves as 
\be
a(t) \sim t^{2/3} \left[1 + C \cos ( 2 m_\phi t +\theta) /t^2 \right] 
\label{a-of-t-phi}
\ee 
 to be compared with that in $R^2$-theory~(\ref{a-of-t-R}).  
 
The energy density of the matter field (of the scalar $\phi$) drops down as $1/t^2$ to be compared with the slow drop-off, as $1/t$, 
of the matter fields in $R^2$ theory. The curvature scalar in this model is proportional to the trace of the energy-momentum tensor of
$\phi$ and monotonically decreases as $1/t^2$, while in $R^2$-theory the curvature behaves as $\cos (mt +\theta)/t$ and is not 
connected with the energy density of the normal matter.

A rather long regime during which the cosmological evolution differs from the standard FLRW cosmology
could lead, in particular,  to modification of high temperature baryogenesis scenarios, to a variation of
the frozen abundances of heavy dark matter particles,
and to necessity of  reconsideration of the formation of primordial black holes.

 Next we consider much larger time, when  $\Gamma t \gg 1$, and study the approach to the usual GR cosmology. 
 GR is recovered when the energy density of matter becomes larger than that of the exponentially decaying scalaron.
We argue, however, that the approach is somewhat delayed. It
takes place not at $\Gamma t \sim 1$, as it may be naively expected,  but at $\Gamma t \sim \ln (m/\Gamma)$. 
}

The paper is organized as follows. In Sec.~\ref{s-cosm-eq} we present and discuss the gravitational equations
of motion modified by the addition of the $R^2$-term into the action. In contrast to some previous papers we included the term
describing particle production as  a source into equation for the energy density evolution~(\ref{dot-rho-pp}). 
These equations are rewritten in a convenient dimensionless form and solved numerically and analytically 
in the next section~\ref{s-initio}. There is an excellent agreement between numerical and analytical results.
The results of this section are obtained in the limit of rather early universe when it was younger than the inverse decay rate of the curvature scalar, $t_U \lesssim 1/\Gamma$. In section~\ref{s-large-gamma-tau} this restiction
is lifted and deep asymptotics of the solution at $\Gamma t \gg 1$ is studied. It is shown that the 
curvature oscillations indeed decay as $\exp (-\Gamma t /2)$ the cosmology returns to the normal
GR one.

\section{Cosmological equations in $R^2$-theory \label{s-cosm-eq}}
 
Let us consider the theory described by the action: 
\be
S_{tot} = -\frac{m_{Pl}^2}{16\pi} \int d^4 x \sqrt{-g} \left(R-\frac{R^2}{6m^2}\right)+S_m\,,
\label{S-R2-tot}
\ee
where  $m$ is a constant parameter with dimension of mass and
$S_m$ is the action of the matter fields.

The modified Einstein equations for theory (\ref{S-R2-tot}) are the following:
\be
 R_{\mn} - \frac{1}{2}g_{\mn} R -
 \frac{1}{3m^2}\left(R_{\mn}-\frac{1}{4}R g_{\mn}+g_{\mn} D^2-  D_\mu D_\nu\right)R
 =\frac{8\pi}{\mpl^2}T_\mn\,, \label{field_eqs}
\ee
where $D^2\equiv g^\mn D_\mu D_\nu$ is the covariant D'Alembert operator. 
The energy-momentum tensor of matter $T_\mn$ is assumed to have the following  diagonal form:
\be
T^\mu_\nu = diag(\rho, -P, -P, -P),
\label{T-mn}
\ee
where $\rho$ is the energy density, $P$ is the pressure of matter.

We assume  that the matter distribution is homogeneous and isotropic 
with the equation of state 
\be
P = w \rho,
\label{eq-state}
\ee
where $w$ is usually a constant parameter. For non-relativistic matter 
$w=0$, for relativistic matter $w=1/3$, and for the vacuum-like state $w=-1$.

The cosmological metric is taken in the standard Friedmann-Robertson-Walker (FRW) form with 
the interval given by
\be
ds^2 = dt^2 - a^2(t)\left[\frac{dr^2}{1-kr^2}+r^2d\vartheta^2+r^2\sin^2\vartheta\,d\varphi^2\right]\,.
\label{FRW}
\ee
In what follows we assume that the  three-dimensional space is flat and thus take $k = 0$.
In this case the curvature scalar $R$ is expressed through the Hubble parameter $H = \dot a/a$ as
\be
R=-6\dot H-12H^2\,.
\label{R-of-H}
\ee

If there are no extra sources of energy created by gravity itself,
the energy-momentum tensor satisfies the covariant conservation condition $D_\mu T^\mu_\nu = 0$, which in 
FRW-metric (\ref{FRW})   has the form: 
\be
\dot\rho = -3H(\rho+P)  = -3H (1+w) \rho\,.
\label{dot-rho}
\ee

Taking the trace of Eq.~(\ref{field_eqs}) yields
\be
D^2 R + m^2 R = - \frac{8 \pi m^2}{m_{Pl}^2} \, T^\mu_\mu.
\label{D2-R}
\ee
The General Relativity  limit should be recovered when $m\rar \infty$.
In this case we expect to obtain the usual algebraic relation between the curvature scalar and the trace of the energy-momentum tensor of matter:
\be
m_{Pl}^2 R_{GR} = - 8\pi T_\mu^\mu\,.
\label{G-limit}
\ee

For homogeneous field, $R=R(t)$, and for the equation of state of matter (\ref{eq-state}) 
equation (\ref{D2-R}) turns into 
\be
\ddot R + 3H\dot R+m^2R = - \frac{8\pi m^2}{\mpl^2}(1 - 3w)\rho \,.
\label{ddot-R}
\ee
This is the Klein-Gordon (KG) type equation for massive scalar field $R$, which is sometimes called ``scalaron''. 
It differs from the usual KG equation by the liquid friction term  $3H \dot R$ with the friction
coefficient $3H$ related to $R$ through Eq.~(\ref{R-of-H}).

This equation does not include the effects of particle production by the curvature 
scalar. It is a good approximation at inflationary epoch, when particle production by $R(t)$ is practically
absent because $R$ is large and friction is large, so $R$ slowly evolves down to zero. At some stage,
when $H$ becomes smaller than $m$, $R$ starts to oscillate efficiently producing particles. 
It commemorates the end of inflation, the heating of the universe, which was 
originally void of matter, and the transition from the accelerated expansion (inflation) to a de-accelerated one.
The latter resembles the usual Friedmann expansion regime but, as we see below, differs in many essential
features.

For the harmonic potential the particle production can be approximately described by an additional 
friction term $\Gamma \dot R$. The effects of particle production for an arbitrary potential in Klein-Gordon
equation are calculated in Ref.~\cite{AD-SH} { in one loop approximation. The results of this work were 
modified for the case of particle production by the curvature scalar in ref.~\cite{ADR-R2}. Generally the one-loop
effects on the particle production lead to non-local in time inegro-differential equation, but in the case of strictly harmonic
oscillations the equation can be reduced to  a simple differential equation with the liquid friction term $\Gamma \dot R$. }
Here, in the case under scrutiny, the potential is harmonic and
we can use the friction term approximation. The particle rate, as calculated in Ref.~\cite{ADR-R2} (see 
also { the earlier works~\cite{YaBZ-AAS,Starobinsky_1982,Vilenkin_1985}),} is equal to:
\be
\Gamma = \frac{m^3}{48 m_{Pl}^2}.
\label{Gamma}
\ee  
Correspondingly  equation (\ref{ddot-R}) acquires an additional friction term and turns into: 
\be
\ddot R + (3H + \Gamma) \dot R+m^2R = - \frac{8\pi m^2}{\mpl^2}(1 - 3w)\rho \,.
\label{ddot-R-Gam}
\ee
Particle production leads also to an emergence of the source term in Eq.~(\ref{dot-rho}):  
\be
\dot\rho = -3H (1+w) \rho + \frac{m R^2_{ampl}}{1152\pi} \,,
\label{dot-rho-pp}
\ee
where $R_{ampl} $ is the amplitude of $R(t)$-oscillations, see Refs.~\cite{Vilenkin_1985,ADR-R2}. 
For simplicity the produced particles are supposed to be
massless scalars, though it is not necessarily  so. 
{ If the produced particles are strictly massless or very light, then the universe would be populated by relativistic matter.
However, if the particle mass is comparable to $m/2$ (but slightly smaller than it), practically the same expression for the scalaron
decay width (\ref{Gamma}) is applicable with the mild phase space suppression factor $\sqrt{ 1 - 4m^2_0/m^2}$, 
where $m_0$ is the mass of the particle
produced in the decay. For example for  $m_0 = 0.4 m$ the suppression is only by 0.6. Higher mass particles may be
produced non-perturbatively, as is described in the papers quoted above eq.~(\ref{Gamma}). 

The state of the cosmological  matter depends not only upon the spectrum of the  decay products but also on the
thermal history of the produced particles.
Depending on that,  the parameter $w$ may be not exactly equal to 0 or 1/3 and the equation of state can 
be not that simple. It may be even impossible to describe it by a constant $w$.
We took the two limiting values $w=0$ and $1/3$ as possible simple examples. Different values of $w$ would not change the presented
results significantly.
Since the spectrum of elementary particles at very high masses is unknown, there is not much sense in doing detailed quantitative
analysis of the state of the cosmological matter, but surely it would be somewhere between the $w=1/3$ and $w=0$ limits.
}

It is convenient to introduce dimensionless time variable and dimensionless functions:
\be
 \tau = t m,\ \ \ H = m h, \ \ \ R = m^2 r, \ \ \ \rho = m^4 y, \ \ \
\Gamma = m \gamma.
\label{dim-less}
\ee
Equations (\ref{R-of-H}), (\ref{ddot-R-Gam}), and (\ref{dot-rho-pp}) now become:
\be
h' + 2h^2 &=&  - r/6, \label{h-prime} \\
r'' + (3h + \gamma) r' + r &=& - 8 \pi \mu^2 (1-3w) y, \label{r-two-prime}\\
y' + 3(1+w)h\,y &=& S[r], \label{y-prime}
\ee
where prime means derivative over $\tau$, $\mu = m/m_{Pl}$, $\gamma = \mu^2/48$, and the source term $S[r]$ is taken as 
\be
S[r] =  \frac{\langle r^2 \rangle} {1152 \pi}.
\label{source}
\ee
The impact of this term on the evolution of the curvature scalar was not properly taken into
account in the { previous} works.
Here  $ \langle r^2\rangle$ means amplitude squared of harmonic oscillations, $r^2_{ampl}$,
 of the dimensionless curvature $r(\tau)$, compare to 
Eq.~\eqref{dot-rho-pp}. However, it is not always true, that $r(\tau)$  oscillates harmonically. In this case we approximate 
$ \langle r^2\rangle$ as $2 (r')^2$ or $(-2 r'' r)$. For harmonic oscillations these expressions averaged over oscillation period
coincide with $r^2_{ampl}$.

 The function  $ \langle r^2 \rangle$ slowly changes with time. 
Strictly speaking such a form for the description of the particle creation is true only during the epoch when $r(\tau)$
is a harmonically oscillating function with slowly varying amplitude. So it is surely inapplicable during inflation. In principle
we can switch on this source only after inflation is over. However, the ultimate result for $y$ (or $\rho$) does not depend on the history of the particle production. The reason for that is the following: 
during inflation the energy density of the normal matter very quickly red-shifted away and we arrive to the moment of the
universe heating with essentially the same, vanishingly small, value of $y$ (or $\rho$). 
In other words, initial condition for the energy density of matter at the onset of the particle production is always $y=0$ ($\rho = 0$). 
Below we show numerically that this is indeed true with 
 very high precision. 

We check compatibility of equations~(\ref{Gamma}) and (\ref{source}) in subsection~\ref{ss-part-prod}, 
where we show
that the particle production through the scalaron decay gives the necessary influx of energy, $S[r]$,
to cosmic plasma.

In what follows we solve the system of equations (\ref{h-prime}) - (\ref{y-prime}) numerically and 
compare the numerical solutions with analytical asymptotic expansion at $\tau \gg 1$. 
The agreement is perfect. This permits to use analytic asymptotic expressions when the numerical
calculations become non-accurate.

\section{Solution {\it ab ovo} to $\gamma \tau \lesssim 1$ \label{s-initio}}

\subsection{Solution at inflationary epoch \label{ss-infl-sol}}

In this section we perform numerical and analytical solutions of Eqs.~(\ref{h-prime}) - (\ref{y-prime}) 
starting from the very beginning, i.e. from the inflationary stage up to high $\tau $ ($\tau \gg 1$), 
but small $\gamma \tau \lesssim 1$. The initial conditions should be chosen in such a way that 
at least 70 e-foldings during inflation are ensured:
\be
N_e = \int_0^{\tau_{inf}} h\,d\tau \geq 70,
\label{h-dt}
\ee
where $\tau_{inf}$ is the moment when inflation terminated.  This can be achieved if the initial value of $r$
is sufficiently large, practically independently on the initial values of $h$ and $y$. 

\begin{figure}[!htbp]
  \centering
  \begin{minipage}[b]{0.45\textwidth}
    \includegraphics[width=\textwidth]{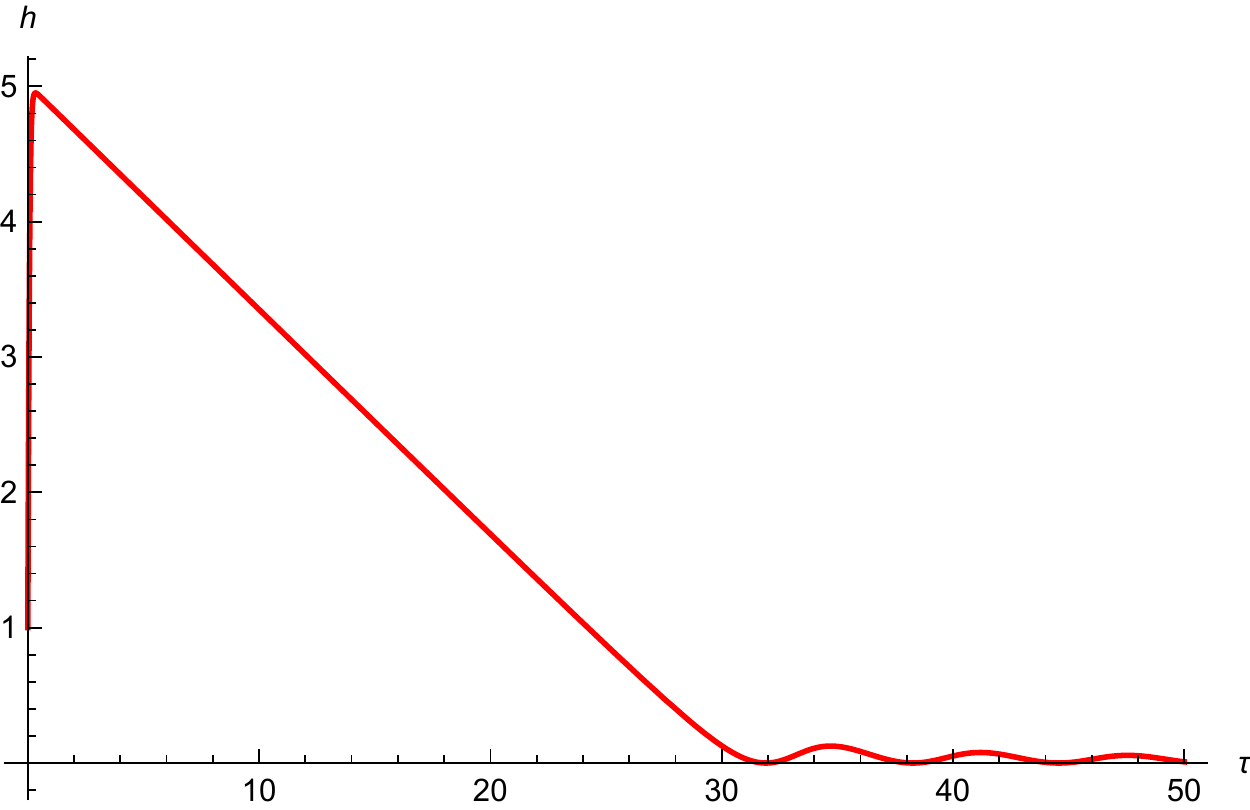}    
  \end{minipage}
  \hspace*{.1cm}
  \begin{minipage}[b]{0.45\textwidth}
    \includegraphics[width=\textwidth]{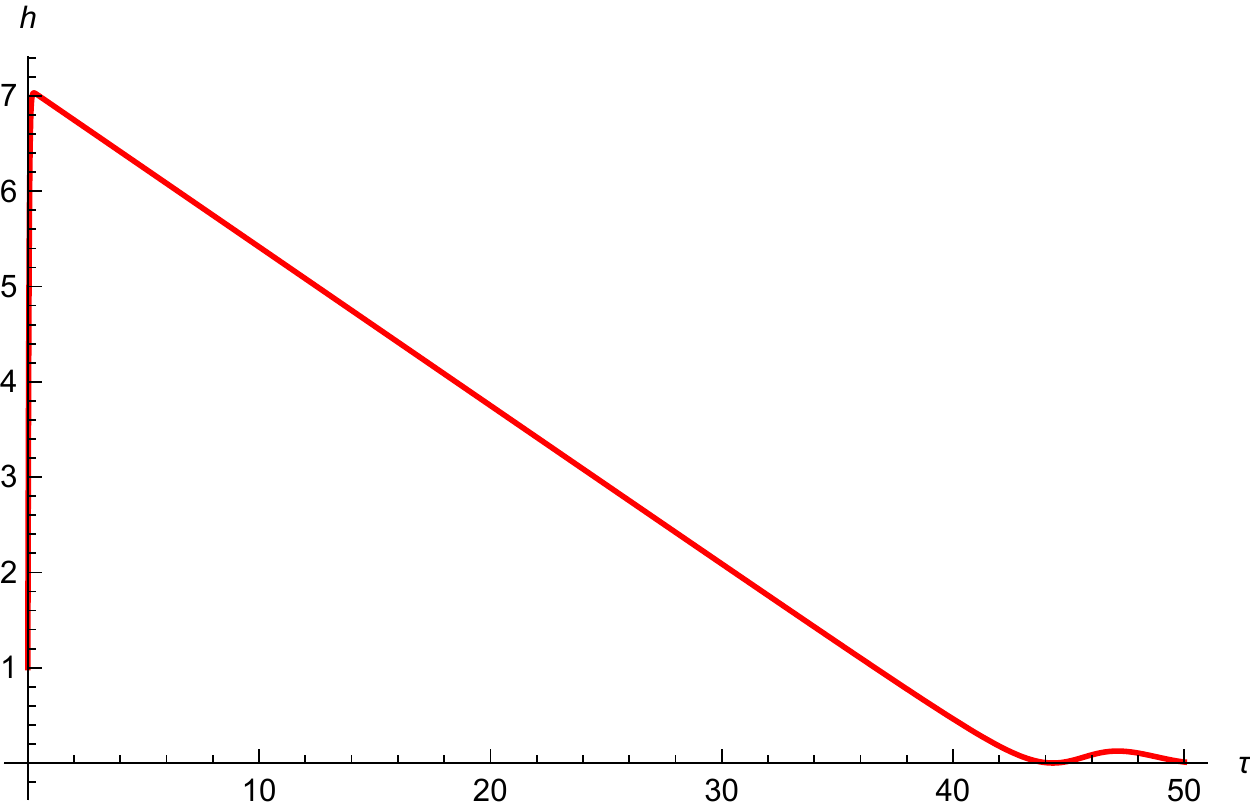}
      \end{minipage}
  \caption{Evolution of $h (\tau)$ at the inflationary stage with the initial values of dimensionless curvature
  $r = 300$ (left) and $600$ (right). Initially $h$ is taken to be zero, $h_{in} = 0$, but it quickly reaches the value given 
  by Eq.~\eqref{h2}, 
  $h(0) = \sqrt{-r_0/12}$.
   The numbers of e-foldings, according to  Eq. (\ref{h-dt}), are respectively 75 and 150.}
  \label{f:h-dt}
 \end{figure}

\begin{figure}[!h]
  \centering
  \begin{minipage}[b]{0.48\textwidth}
    \includegraphics[width=\textwidth]{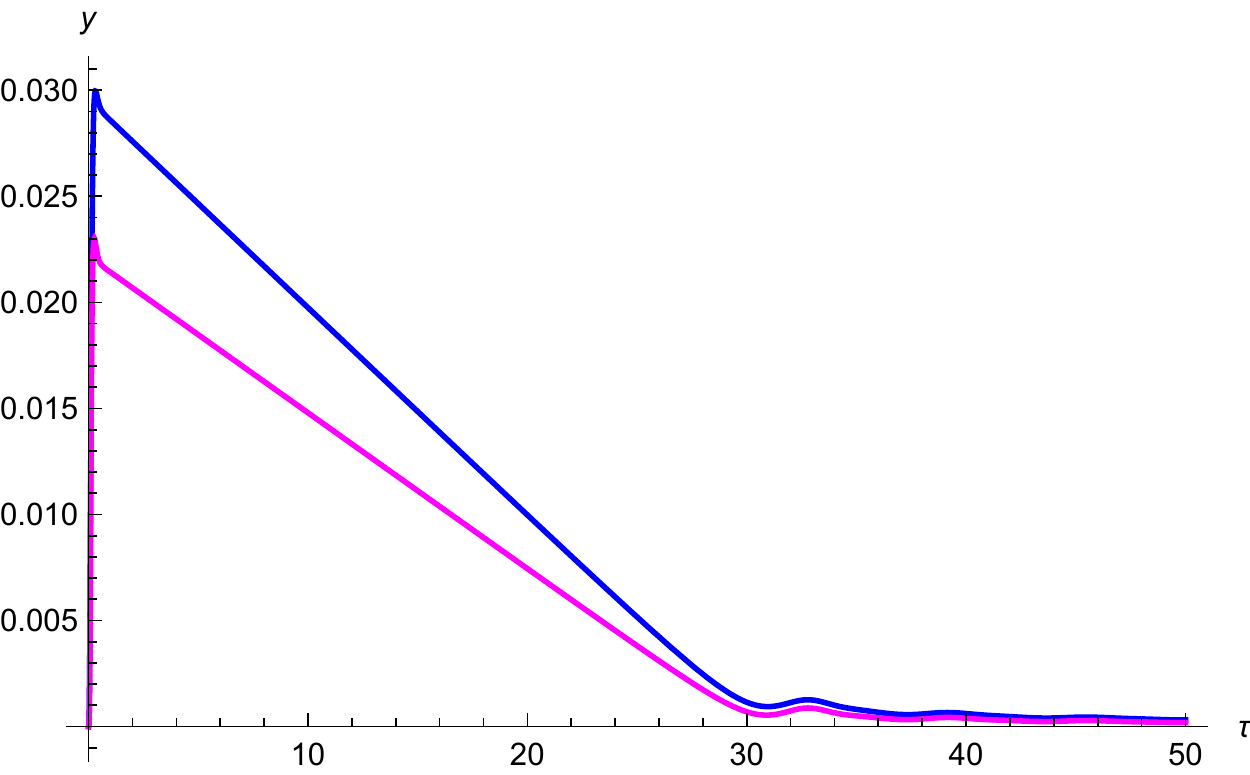}
      \end{minipage}
  \begin{minipage}[b]{0.48\textwidth}
    \includegraphics[width=\textwidth]{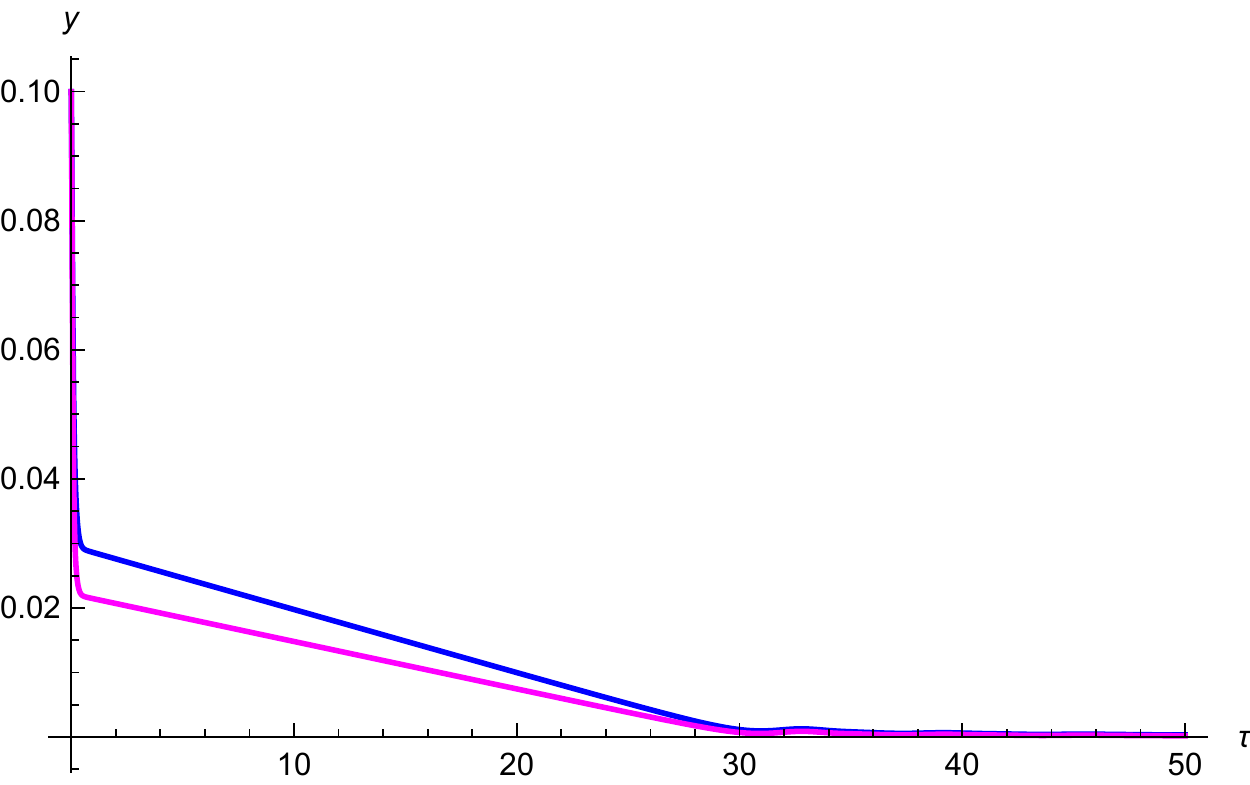}
\end{minipage}
  \caption{Evolution of the dimensionless energy density of matter during inflation for $w = 0$ (blue) and $w=1/3$ (magenta).
   Left panel: initially $y_{in}= 0$ and  right panel: $y_{in} = 0.1$. 
  The initial fast rise of $\rho$ from zero in the left panel during
  short time is generated by the  $ S[r] $-term 
  (\ref{source}) taken as  $ S[r] = (r')^2 /288 \pi $. The results are not sensitive to the form $S[r]$ because 
  at inflation $y(\tau)$ quickly vanishes anyhow.
 }
  \label{f:rho-inf}
 \end{figure}

{ Following refs.~\cite{Starobinsky_1980,R2-rev} (see also the subsequent work~\cite{koshelev}),}
we can roughly estimate the duration of inflation neglecting higher derivatives in 
Eqs.~(\ref{h-prime}) and (\ref{r-two-prime}) and assume that the energy density of the usual matter 
vanishes ($y=0$) and $\gamma $ is negligibly small. The latter is naturally achieved if $m <m_{Pl}$. 
So we arrive to the simplified set of equations:
\be
h^2 &=& - r/12, \label{h2}\\
3 h r'  &=& - r  . \label{h-r-prime}
\ee
These equations are solved as:
\be
\sqrt{-r(\tau)} =  \sqrt{-r_0} - \tau/\sqrt 3,
\label{sqrt-r}
\ee
where $r_0$ is the initial value of $r$ at $\tau = 0$. According to Eq. (\ref{h2}), the Hubble parameter  
behaves as $h(\tau)=(\sqrt{-3r_0} - \tau)/6$. The duration of inflation is roughly determined by the condition 
$h=0$, i.e. $\tau_{inf} = \sqrt{-3r_0}$. The number of e-folding is equal to the area of the triangle below 
the line $h(\tau)$, thus $N_e \approx r_0 / 4$. It is in excellent agreement with numerical solutions of
Eqs.~(\ref{h-prime}-\ref{y-prime}) depicted in Fig.~\ref{f:h-dt}. { This demonstrates high precision of the slow roll
approximation and weak impact of particle production at (quasi)inflationary stage.}
Numerical results are neither sensitive to the initial values of the Hubble parameter and of 
the energy density of { usual} matter, nor to the parameter $w$,
because at inflation any preexisting matter density is quickly washed out. This statement is illustrated by 
Fig.~\ref{f:rho-inf}. 

\begin{figure}[!htbp]
  \centering
    \includegraphics[width=0.5\textwidth]{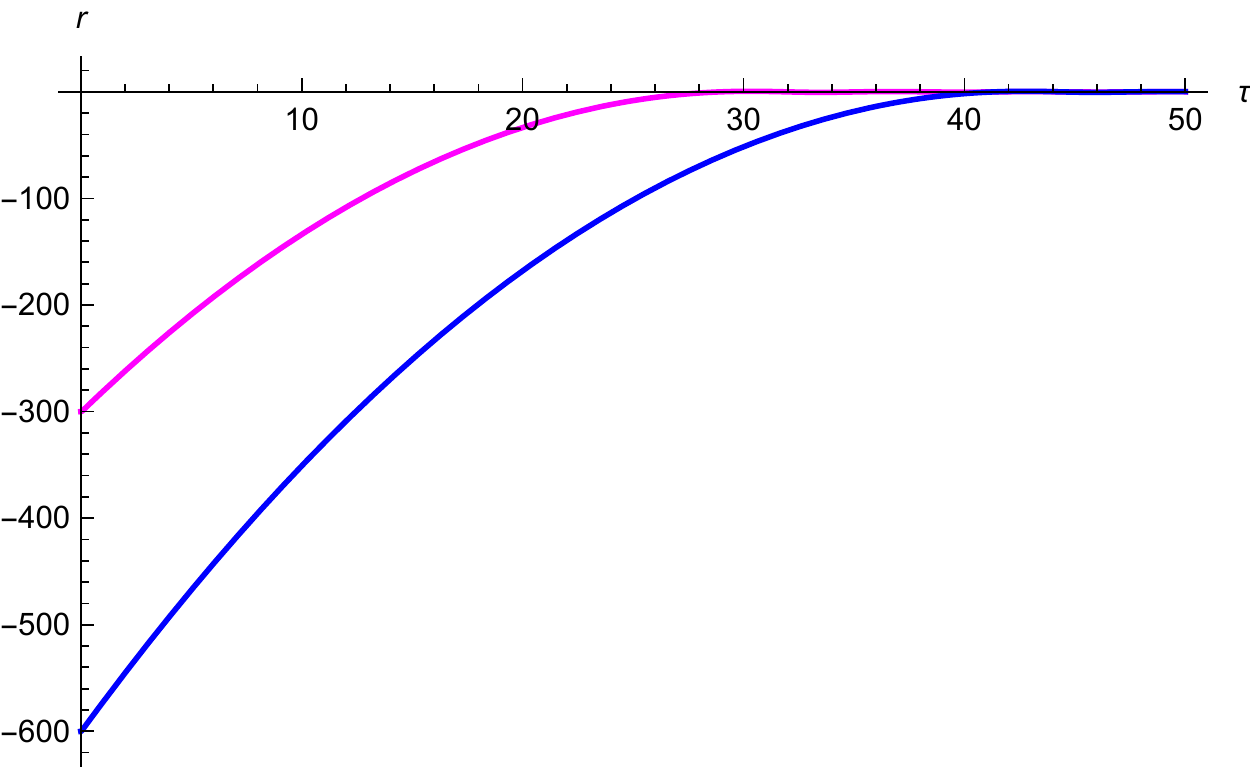}
  \begin{minipage}[b]{0.48\textwidth}
    \includegraphics[width=\textwidth]{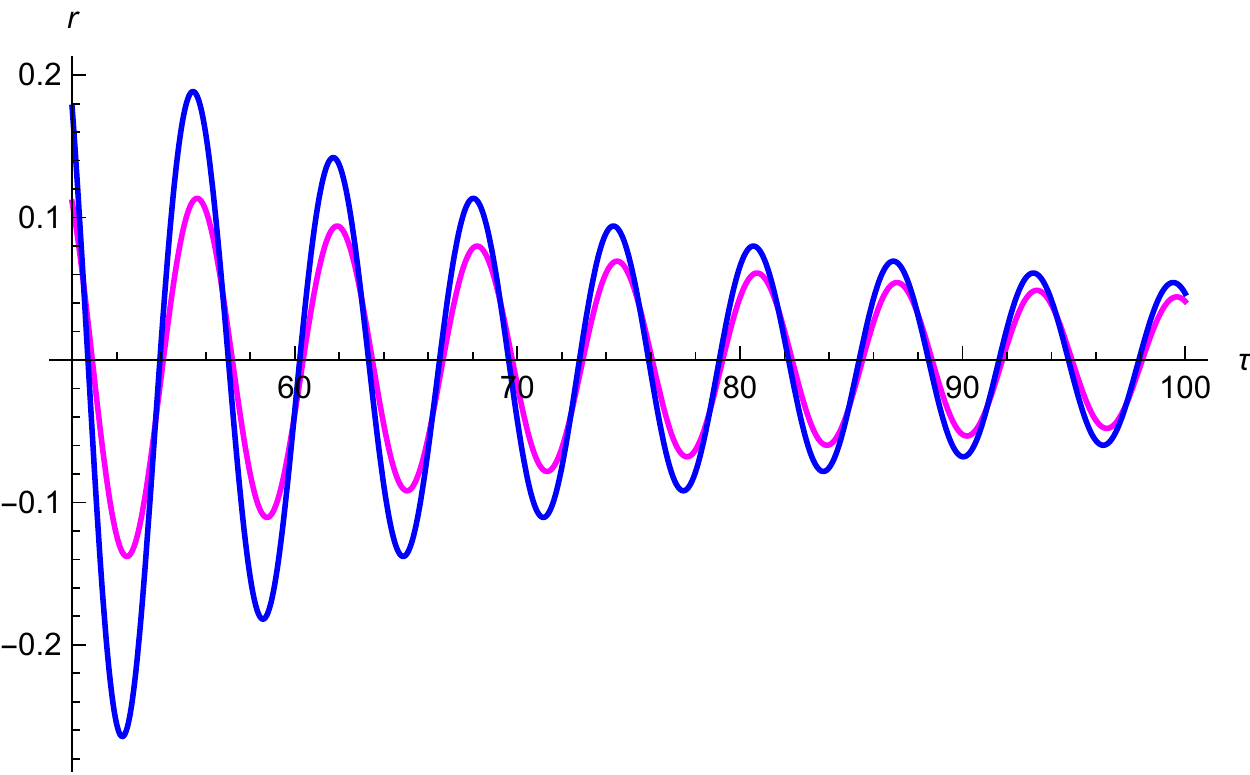}
\end{minipage}
 \caption{Evolution of the dimensionless curvature scalar for $r_{in}=-300$ (magenta)
and  $r_{in}=-600$ (blue). Left panel: shows evolution during inflation and right panel: shows evolution after the end of inflation when curvature scalar starts to oscillate.
}
  \label{f:r36}
 \end{figure}

The evolution of the dimensionless curvature scalar, $r$, during inflation is presented in Fig.~\ref{f:r36}.

\subsection{Numerical solutions at post-inflationary epoch} \label{ss-num-sol-MD}

The behavior of $R$, $H$ and $\rho$, or dimensionless quantities $r$ , $h$, and $y$ is
drastically different at  the vacuum-like dominated stage  (inflation) and during matter dominated (MD) stage,
which followed the inflationary epoch.  Now we will find the laws of evolution of $r(\tau)$, $h(\tau)$, and $y(\tau)$
after inflation till $\gamma \tau \sim 1$. The numerical solutions will be presented from the end of inflation
to large $\tau \gg 1 $, but not too large because the numerical procedure for huge $\tau \sim 1/\gamma$
becomes unstable. However, we can find pretty accurate analytical solution, asymptotically valid at any large
$\tau$  up to $\tau \sim 1/\gamma$. Very good agreement between numerical and analytical
solutions at large but not huge $\tau$ allows to trust asymptotic analytical solution at huge $\tau$. 

\begin{figure}[!htbp]
  \centering
  \begin{minipage}[b]{0.48\textwidth}
    \includegraphics[width=\textwidth]{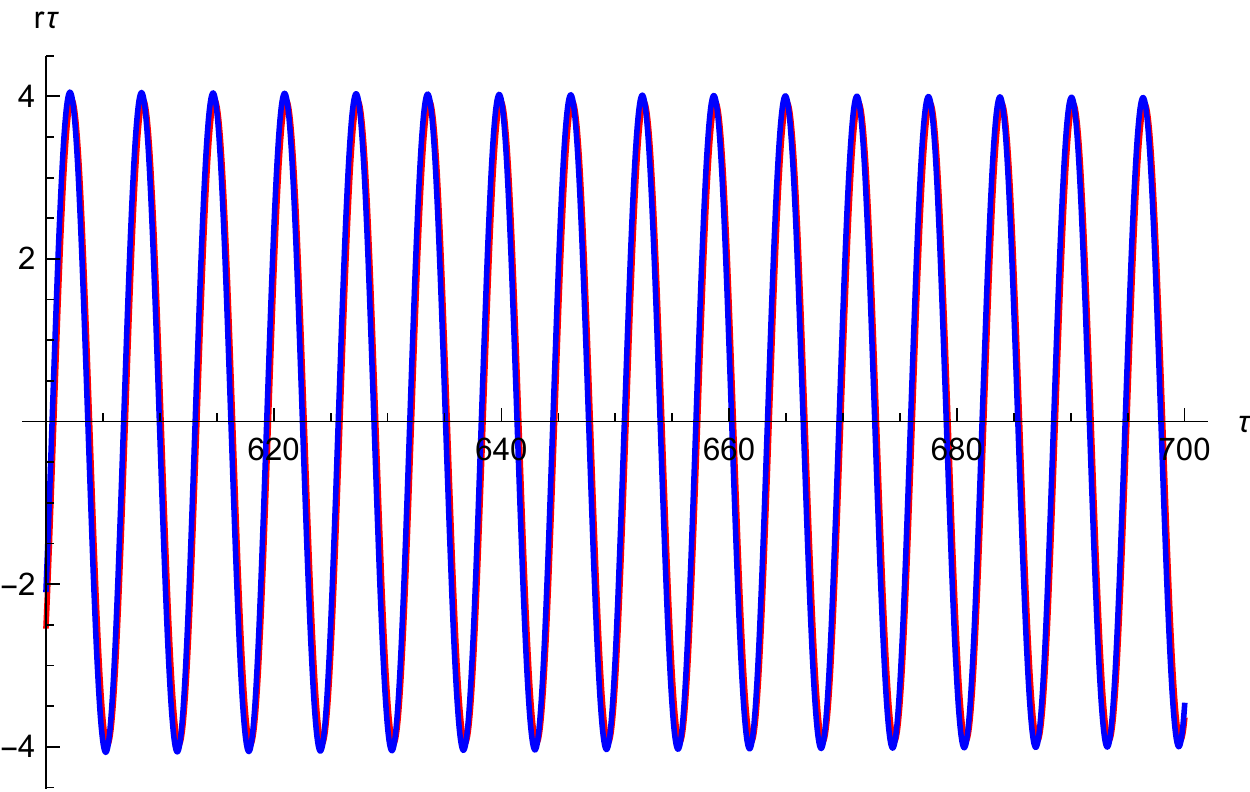}
      \end{minipage}
  \hfill
  \begin{minipage}[b]{0.48\textwidth}
    \includegraphics[width=\textwidth]{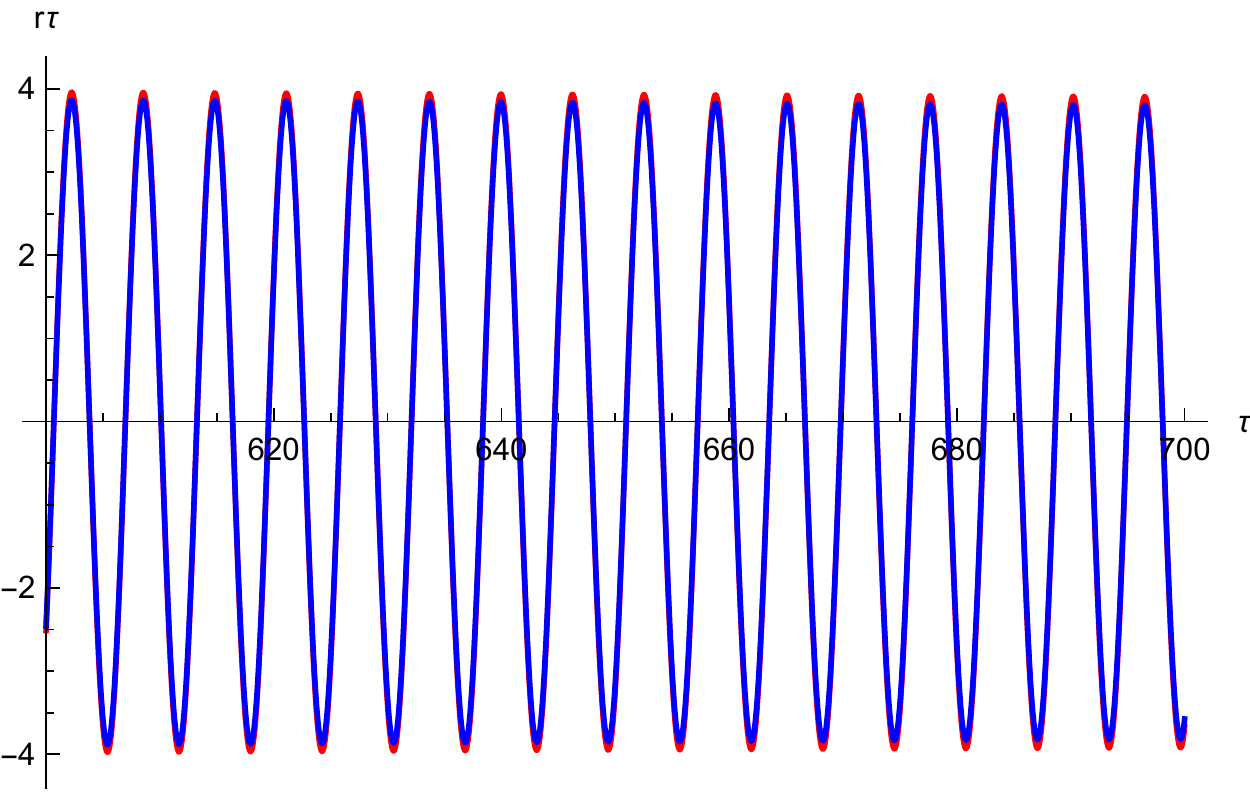}
\end{minipage}
  \caption{Evolution of the curvature scalar $ \tau r(\tau)$
  in post-infationary epoch. 
 Left panel (w=1/3): initially $r_{in}= -300$ (red),  $r_{in} = -600$ (blue). There is absolutely no difference between the curves. 
  Right panel ($r_{in}= -300$):  $w=1/3$ (red) and $w= 0$ (blue). The difference is minuscule.
   The source term (\ref{source}) here is  taken as $ S[r] = (r')^2 /1152 \pi $. The results are not sensitive to its form. }
  \label{f:r-postinf}
 \end{figure}

\begin{figure}[!htbp]
  \centering
    \includegraphics[width=0.52\textwidth]{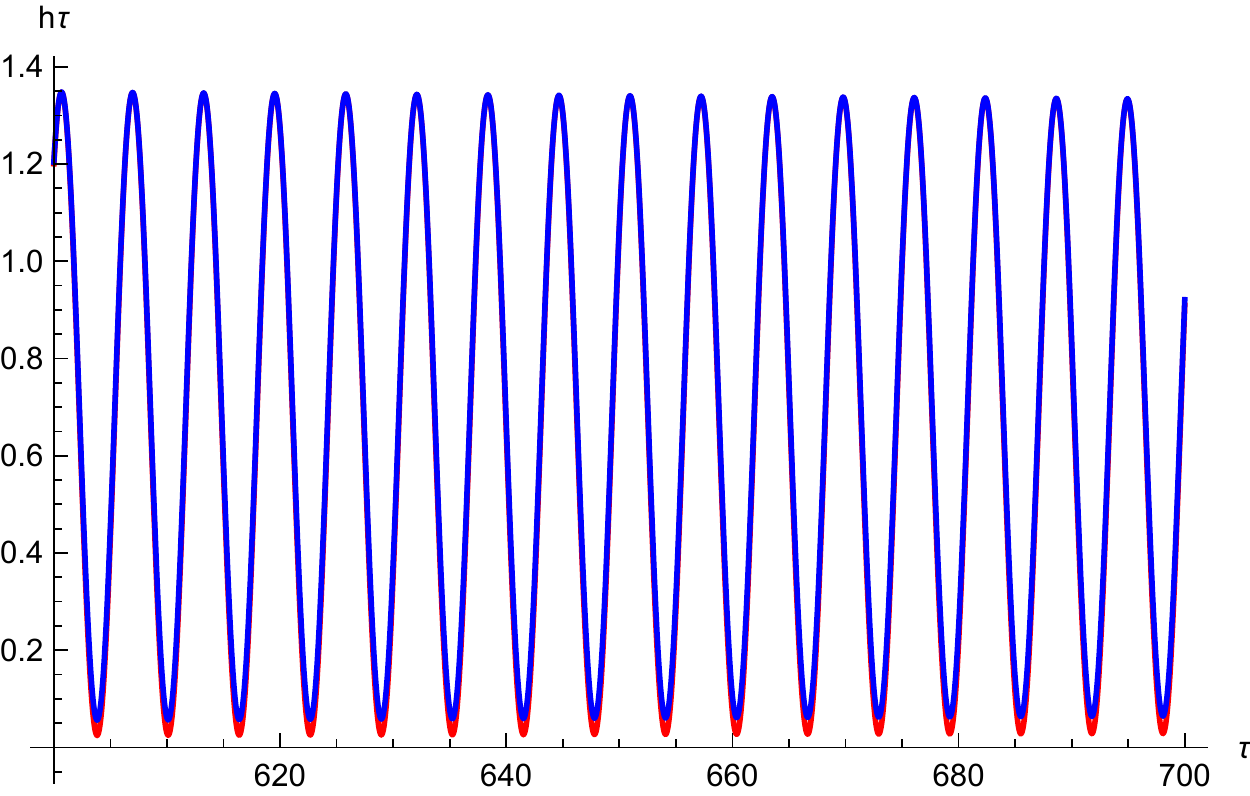}
 \caption{
Evolution of the Hubble parameter, $h \tau$, in post-inflationary epoch for $w=1/3$ (red) 
and $w = 0$ (blue).   
 }
  \label{f:hw}
 \end{figure}

\begin{figure}[!htbp]
  \centering
  \begin{minipage}[b]{0.48\textwidth}
    \includegraphics[width=\textwidth]{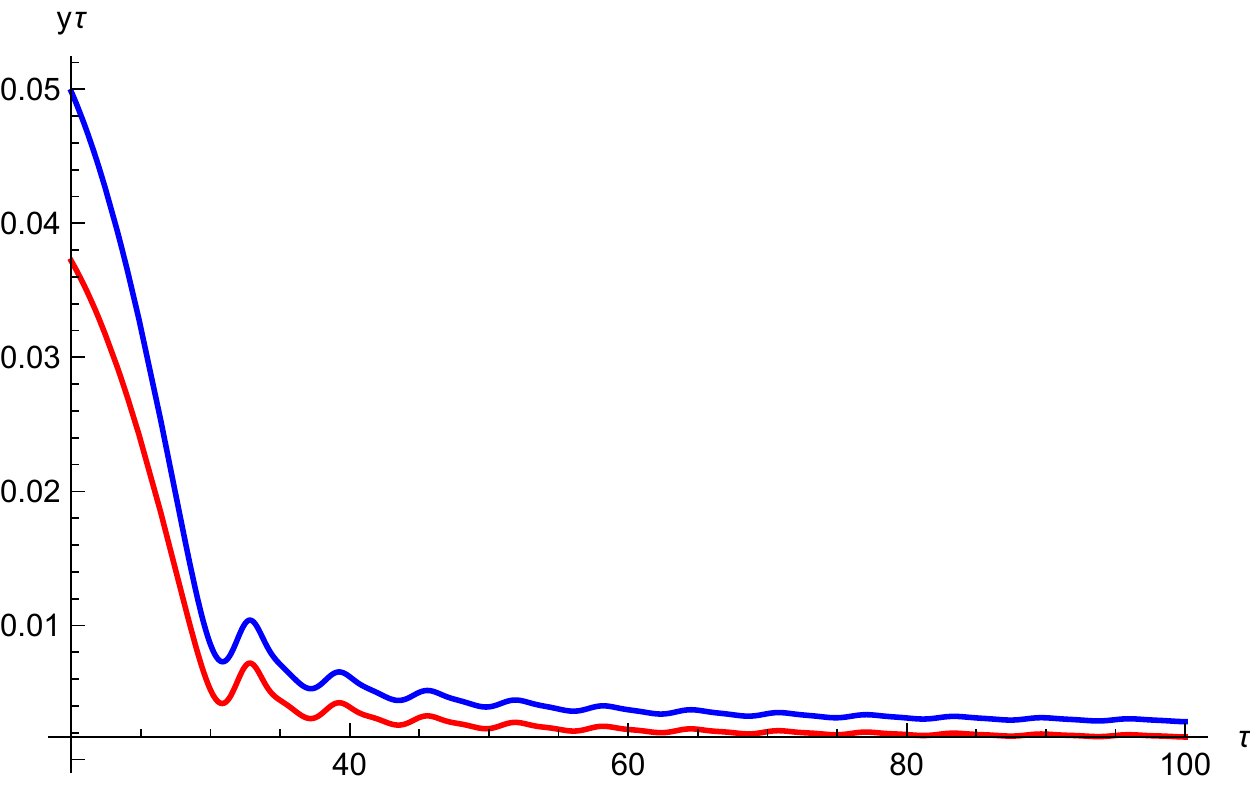}
      \end{minipage}
  \begin{minipage}[b]{0.48\textwidth}
    \includegraphics[width=\textwidth]{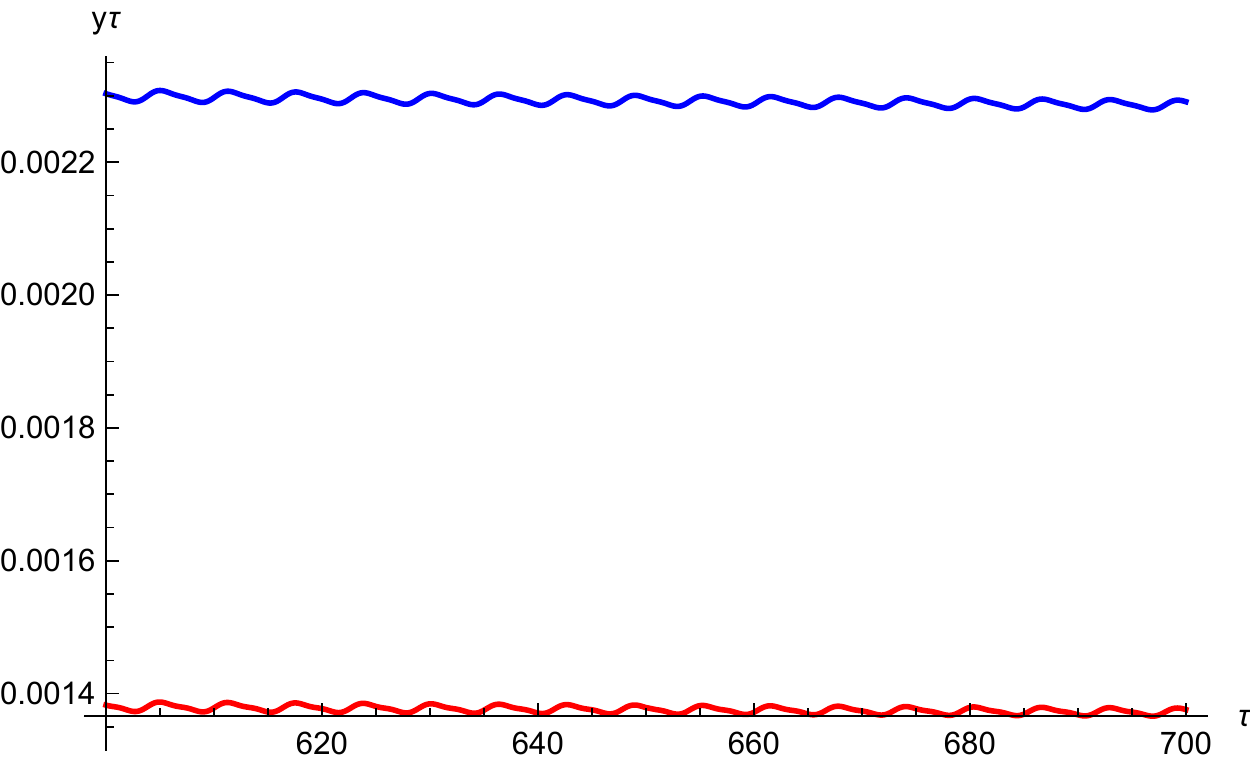}
\end{minipage}
  \caption{
 Evolution of the energy density of matter $y \tau$ at small $\tau$ (left) and at large
 $\tau$ (right). Parameter $w=1/3$ (red) and $w = 0$ (blue). 
  }
  \label{f:rho-w}
 \end{figure}

We solved numerically the system of equations (\ref{h-prime})-(\ref{y-prime}). The results are depicted in 
Figs.~\ref{f:r-postinf}-\ref{f:rho-w}. In all the figures we take very large $\mu= m/m_{Pl} = 0.1$. 
In Fig.~\ref{f:r-postinf} the dimensionless curvature is presented for different initial values $r_{in} = -300$ and
$r_{in} = -600$ and for different equations of stateswith $w= 1/3$ (relativistic matter) and $w=0$ (nonrelativistic matter). 
The dimensionless time is large,  $\tau \gg 1$, but still $\gamma \tau < 1$. 

The contribution of particle production to $\rho$ (or $y$) is approximated as 
$ (r' )^2/(576 \pi) $, see Eqs. (\ref{y-prime}), \eqref{source}.
Here we have taken into account the factor 2, appearing because the average value of $\sin^2 \tau = 1/2$.
It is interesting that the amplitude $r_{amp} \tau \rightarrow const $.
We see that for large $\tau$ the result  does not depend upon the initial value of $r$ and very weakly depends 
on $w$.

In Fig.~\ref{f:hw} the evolution of the dimensionless Hubble parameter is presented for $w=1/3$ (red)
and $w =0$ (blue). The dependence on $w$ is very weak, except for small values of $h$ when it approaches 
zero. If $h$ is very close to zero the numerical solution may become unstable because at negative $h$
expansion turns into contraction.

In Fig.~\ref{f:rho-w} the energy density of matter as a function of time is presented for
$w = 1/3 $ (red) and $w=0$ (blue). The magnitude of $\rho$ for these two values of $w$
are noticeably different in contrast to other relevant quantities, $r$ and $h$, which very weakly
depend upon $w$. It is interesting that the product $y \tau $ tends to a constant value with rising
$\tau$ till $\gamma \tau$ remains small. We have checked this statement up to $\tau = 5000$ for $\mu = 0.01$.
This behavior much differs from the matter density evolution in the standard cosmology, when
$\rho \sim 1/t^2$.

We see that the numerical solutions have very simple form for large $\tau$, namely $r$ oscillates 
with the amplitude decreasing as $1/\tau$ around zero, while $h$ also oscillates almost touching zero
with the amplitude also decreasing as $1/\tau$ around some constant value close to 2/3. Such a simple 
behavior hints that there should be simple analytic expressions for the solutions at large $\tau$, which are
derived in the next subsection.

\subsection{Asymptotic behavior of the solution at $\tau \gg 1$ and $w=1/3$}\label{ss-asympt}

Let us start first with a simpler case of $w =1/3$. In this case the system of equations 
(\ref{h-prime}-\ref{y-prime}) turns into 
\be
h' + 2h^2 &=&  - r/6, \label{h-prime-w} \\
r'' + 3h  r' + r &=&0, \label{r-two-prime-w}\\
y' +4 h\,y &=& \frac{<r^2>}{1152 \pi} , \label{y-prime-w}
\ee
so the system is effectively reduced to two equations for $h$ and $r$, while the equation for $y$ can be solved if
$h$ and $r$ are known. The equation for $y$ can be solved either numerically or in terms of quadratures. 
In Eq. (\ref{r-two-prime-w}) we have neglected 
$\gamma $ in comparison with $h$, since by assumption we confine ourselves to the limit 
$\gamma \tau \lesssim 1 $. The case 
$ \gamma \tau \gtrsim 1$ is considered in subsection \ref{ss-asympt-0}.

Stimulated by the numerical solution we search for the asymptotic expansion of  $h$ and $r$  at $\tau \gg 1$ in the form: 
\be
r&=&\frac{r_1 \cos(\tau + \theta_r)}{\tau} + \frac{r_2}{\tau^2}, \label{r-sol1} \\
h&=&\frac{h_0+h_1 \sin(\tau + \theta_h)}{\tau}. \label{h-sol1}
\ee
Here $r_j$ and $h_j$ are some constant coefficients to be calculated from Eqs.(\ref{h-prime-w}) and 
(\ref{r-two-prime-w}), while the constant phases $\theta_j$ are determined through the initial conditions and 
will be adjusted by the best fit of the asymptotic solution to the numerical one.

Substituting expressions (\ref{r-sol1}) and (\ref{h-sol1}) into the r.h.s. of Eq. (\ref{h-prime-w}) we  obtain:
\be
h'&=&-\frac{2[h_0+h_1 \sin(\tau + \theta_h)]^2}{\tau ^2} -\frac{1}{6} \left(\frac{r_1 \cos(\tau + \theta_r)}{\tau}+ \frac{r_2}{\tau^2}\right) ,
\label{h-prime1-1}
\ee
where $h'$ is found by the differentiation of (\ref{h-sol1}):
\be
h'&=& \frac{h_1 \cos(\tau + \theta_h)}{\tau} - \left[\frac{h_0+h_1 \sin(\tau + \theta_h)}{\tau^2}\right] \label{h-prime-3}
\ee

Comparison of the $1/\tau$ and $1/\tau^2$ leads  respectively to the equations:
\be
&&h_1 = -{r_1}/{6}  ,\,\,\,\,  
 \theta_h = \theta_r \equiv \theta, 
 \label{thetahr} \\
&& {h_0+h_1\sin(\tau+\theta_h)} = { 2\left [h_0+h_1 \sin(\tau + \theta_h) \right]^2}+{r_2}/{6} 
\label{h0-h1-r2}
\ee
Neglecting the oscillating $sine$-terms, which vanish on the average, and taking the average of 
$\sin^2(\tau+\theta)= 1/2$, we find:
\be 
h_0=2h_0^2 + h_1^2+ {r_2}/{6} \label{h0eqn-1}.
\ee

Similarly, we explore Eq.~(\ref{r-two-prime-w}) for $r$ and to this end we need to find expressions 
for $r'$, and $r''$:
\be
r'&=& -\frac{r_1 \sin(\tau + \theta_r)}{\tau}- \frac{r_1 \cos(\tau+\theta_r)}{\tau^2} - \frac{2r_2}{\tau^3}, \label{rprime-sol} \\
r''&=&-\frac{r_1\cos(\tau + \theta_r)}{\tau} + \frac{2r_1\sin(\tau+\theta_r)}{\tau^2} + 
\frac{2r_1 \cos(\tau+\theta_r)}{\tau^3}+ \frac{6r_2}{\tau^4}. \label{r2prime-sol}
\ee

Now, substituting the expressions for $r$, $r'$, $r''$, and $h$  into Eq. (\ref{r-two-prime-w}) we arrive at:
\be
&&-\frac{r_1\cos(\tau + \theta)}{\tau} + \frac{2r_1\sin(\tau+\theta)}{\tau^2} + 
\frac{2r_1 \cos(\tau+\theta)}{\tau^3}+ \frac{6r_2}{\tau^4} \nonumber \\
&&- 3 \left[\frac{r_1 \sin(\tau+\theta)}{\tau} + \frac{r_1 \cos(\tau+\theta)}{\tau^2} +
\frac{2r_2}{\tau^3} \right] \left(\frac{h_0+h_1 \sin(\tau + \theta)}{\tau} \right) \nonumber \\
&&+ \frac{r_1 \cos( \tau + \theta)}{\tau} + \frac{r_2}{\tau^2} = 0
\label{r-two-prime-corr}
\ee
The leading terms proportional to $\tau^{-1}$ neatly cancel out. The terms of the order of $\tau^{-2}$ leads to:
\be
h_0 = 2/3,\,\,\,\,\, {\rm and}\,\,\,\,  r_2 =3 h_1 r_1/2.
\label{h0-r2}
\ee
Here, as usually, we  have taken $\sin^2 (\tau+\theta) = 1/2$. Now using Eqs. (\ref{thetahr}) and (\ref{h0eqn-1}), we find:
\be 
h_1 = 2/3\,\,\,\,{\rm and}\,\,\,\, r_1=r_2 = -4,
\label{h1-r1-r2}
\ee
so finally:
\be
&&h= \frac{2}{3\tau} \left[1+ \sin(\tau + \theta )\right] \label{hsol}, \\
&&{r=-\frac{4\cos(\tau+ \theta)}{\tau} - \frac{4}{\tau^2}} . \label{rsol}
\ee
{ As we have already mentioned in the Introduction, these solutions follow from the derived in ref.~\cite{Starobinsky_1980}
expression (\ref{a-of-t-R}) for the evolution of the cosmological scale factor at large $mt$.  The effect of particle production, justly
omitted in ~\cite{Starobinsky_1980}, is shown to be negligible at this stage.}

\begin{figure}[!htbp]
  \centering
  \begin{minipage}[b]{0.48\textwidth}
    \includegraphics[width=\textwidth]{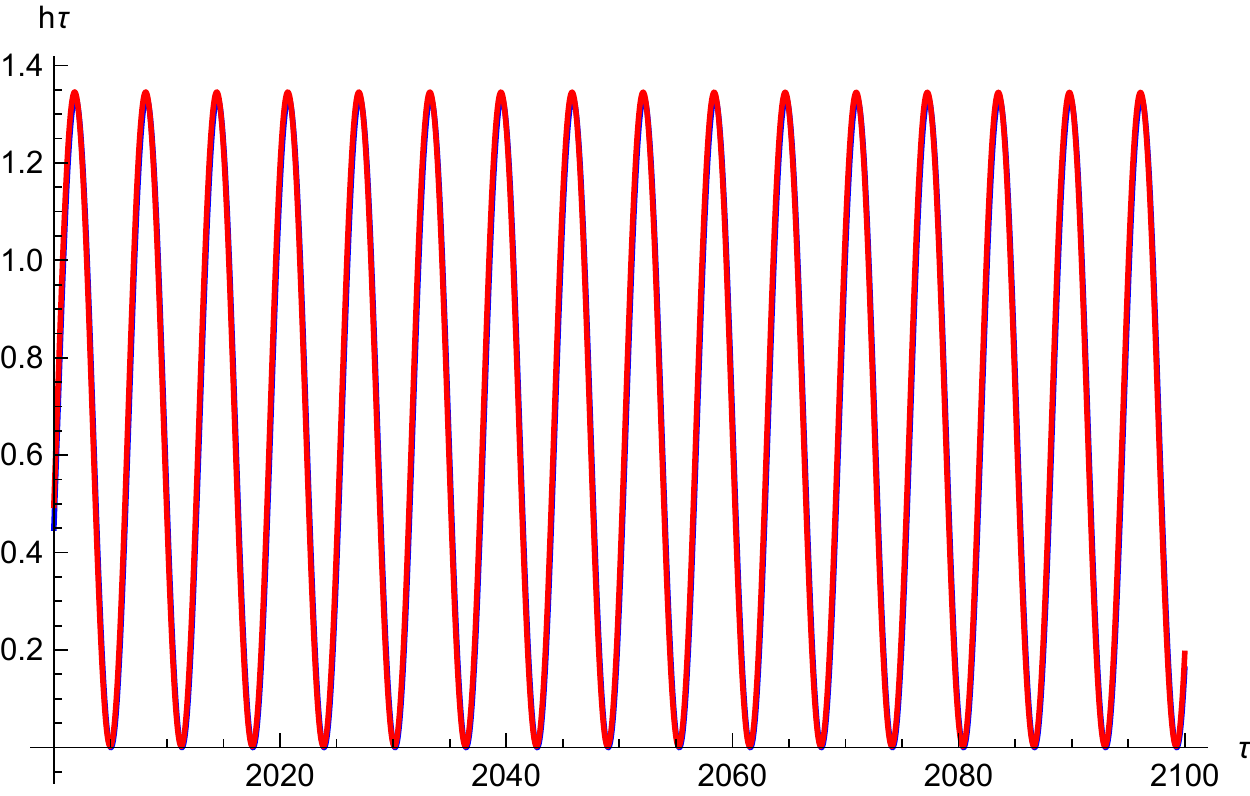}
      \end{minipage}
  \hfill
  \begin{minipage}[b]{0.48\textwidth}
    \includegraphics[width=\textwidth]{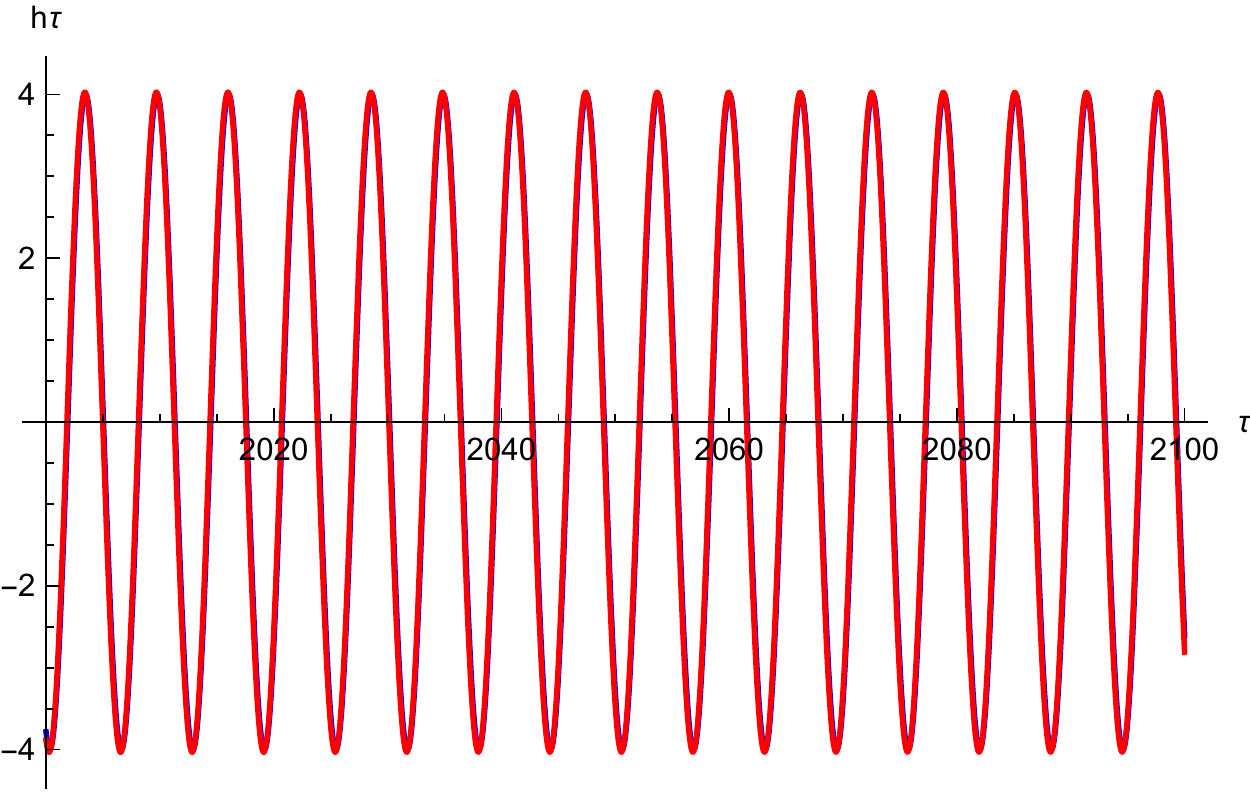}
\end{minipage}
  \caption{
Left panel: comparison of numerical solution for $h \tau$ (red) with analytic estimate (\ref{hsol}) (blue). 
Right panel: the same for numerically calculated $r \tau $ with analytic result (\ref{rsol}). 
 The difference between the red and blue curves is not observable.
 }
  \label{f:h-r-cmpr}
 \end{figure}

In Figs.~\ref{f:h-r-cmpr} results of numerical calculations for $r$ and $h$ are compared with the 
analytic estimates (\ref{hsol}) and (\ref{rsol}) respectively. The phase $\theta$ was adjusted "by hand" 
as $\theta = -2.9\pi/4$, equally  for $h$ and $r$. The value of the phase depends upon the initial 
conditions for $h$ and $r$ prior to inflation.

Now let us turn to solution of Eq. (\ref{y-prime-w}), assuming that $h$ and $r$   have asymptotic 
forms~({\ref{hsol}) and (\ref{rsol}).  Let us first mention that in the r.h.s. of  Eq.~\eqref{y-prime-w}  $\langle r^2 \rangle$
can be understood as the square of the amplitude of the harmonic oscillations of $r$, i.e.  
$\langle r^2 \rangle = 16 /\tau^2 $.
Eq.~(\ref{y-prime-w}) can be analytically integrated as:
\be 
y(\tau) = \frac{1}{72\pi} \,\int_{\tau_0}^\tau \frac{d\tau_2}{\tau_2^2}\,\exp \left[ - 4 \int_{\tau_2}^\tau d\tau_1 h (\tau_1) \right] ,
\label{y-asympt-1/3}
\ee
where $\tau_0 \ll \tau$ is some initial value of the dimensionless time. The asymptotic result weakly depends upon $\tau_0$.

Taking $h(\tau)$ from Eq.~\eqref{hsol} we can partly perform integration over $d\tau_1$ as 
\be 
\int_{\tau_2}^\tau d\tau_1 h (\tau_1)  = \frac{2}{3} \ln \frac{\tau}{\tau_2} + \int_{\tau_2}^\tau \frac{d\tau_1}{\tau_1} \, \sin(\tau_1 + \theta)
\label{int-dtau1}
\ee 

 It is convenient to introduce new integration variables:
\be
\eta_1 = \tau_1/\tau ,\,\,\,\,\, \eta_2 = \tau_2/\tau.
\label{eta-of-tau}
\ee
In terms of these variables we lastly obtain:
\be 
y(\tau) = \frac{1}{72\pi \tau} \,\int_{\eta_0}^1 {d\eta_2}\,{\eta_2^{2/3}}\,\exp \left[ - \frac{8}{3} \int_{\eta_2}^1 \frac{d\eta_1}{\eta_1} \,
\sin \left(\tau \eta_1 +  \theta \right)\right] .
\label{y-asympt-1/3-2}
\ee
As we see below, the integral in the exponent is small, so the exponential factor in expression 
(\ref{y-asympt-1/3-2}) is close to unity
and thus the dominant asymptotic term is $y (\tau) \sim 1/(120\pi \tau) $. Higher order  oscillating corrections 
we estimate as follows. To calculate the asymptotic behavior of the integral
\be
I =\int_{\eta_2}^1 \frac{d\eta_1}{\eta_1} \, \sin \left(\tau \eta_1 +  \theta \right)
 \label{exp-int}
 \ee
 at large $\tau$ we present the oscillating factor as
 \be
  \sin \left(\tau \eta_1 +  \theta \right) = \frac{1}{2i} \left[ e^{i (\tau \eta_1 +  \theta)} - e^{-i (\tau \eta_1 +  \theta)}  \right].
 \label{sin-exp}
 \ee
 The integral over $\eta_1$ along the real axis from $\eta_2$ to 1 can be reduced to two integrals over 
 $\zeta$ from 0 to $\infty$
 along  $\eta_1 = \eta_2 \pm i\zeta$ and $\eta_1 = 1 \pm i\zeta$. The signs in front of $i\zeta$ are 
 chosen so that the corresponding exponent in Eq.~\eqref{sin-exp} vanishes at infinity.
Finally we obtain:
 \be
 I = \int_0^\infty d\zeta e^{-\tau \zeta}   \left[ \frac{\eta_2 \cos \left(\tau \eta_2 +\theta \right) + 
 \zeta \sin \left(\tau \eta_2 + \theta \right)}{\eta_2^2 + \zeta^2}  -
 \frac{ \cos \left(\tau +\theta \right) + \zeta \sin \left( \tau +\theta \right)}{1 + \zeta^2}  
 \right] .
 \label{I}
 \ee
 The  effective value of $\zeta$ in this integral is evidently $\sim 1/\tau$, thus 
  $I$ is inversely proportional to $\tau$ in the leading order. At large $\tau$ it is much smaller than 
  unity, so we can expand the exponential function $\exp (-8 I/3)$, see Eq.~\eqref{y-asympt-1/3-2}, 
  up to the first order and obtain:
\be 
y_{1/3} = \frac{1}{120\pi \tau} +\frac{1}{45\pi} \frac{ \cos \left( \tau + \theta\right)}{\tau^2} - 
\frac{1}{27\pi \tau^2} \,\int_\epsilon^1\frac{d \eta_2}{\eta_2^{1/3} } 
\,\cos \left( \tau \eta_2 + \theta  \right) ,
\label{y-fin-asym}
\ee 
where the subindex (1/3) indicates that $w = 1/3$ and
$\epsilon = \tau_0 /\tau \ll 1$. The last integral is proportional to $1/\tau^{2/3}$ and is subdominant. 
We neglect it in what follows.
 
 In Fig.~\ref{f:y-compar} the dimensionless energy density $120\pi y(\tau)$ is presented as a result of numerical calculation of the 
 integral~(\ref{y-asympt-1/3}) or equivalently (\ref{y-asympt-1/3-2})
 (blue), which is the exact solution of the differential equation~(\ref{y-prime-w}). 
 It is compared with the analytic asymptotic expansion (\ref{y-fin-asym})
 of the same integral (\ref{y-asympt-1/3}) (red). 
 The calculations are done
 for mildly and very large $\tau$. One can see that the agreement is perfect.  However, the numerical calculations of 
 integral (\ref{y-asympt-1/3-2}) take  rather long time.

 \begin{figure}[!htbp]
  \centering
  \begin{minipage}[b]{0.4\textwidth}
    \includegraphics[width=\textwidth]{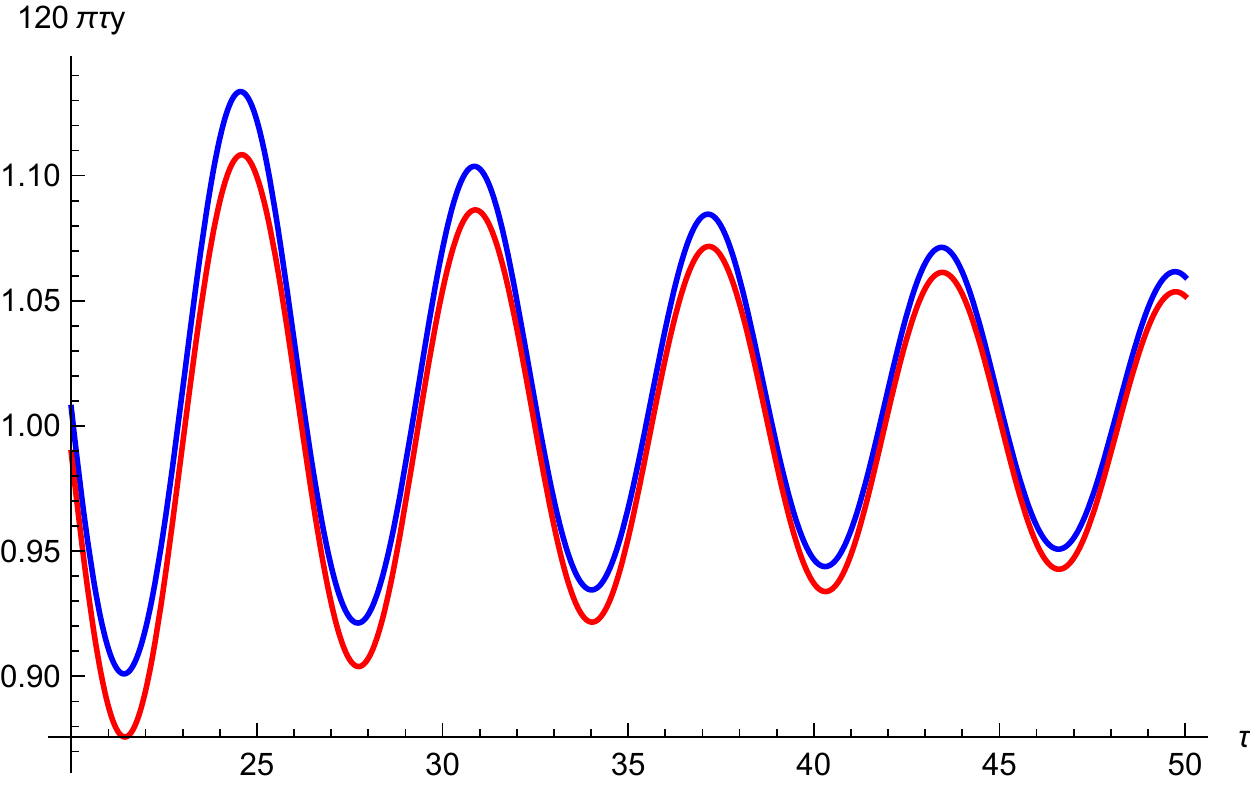}
      \end{minipage}
\hspace*{.6cm}
  \begin{minipage}[b]{0.4\textwidth}
    \includegraphics[width=\textwidth]{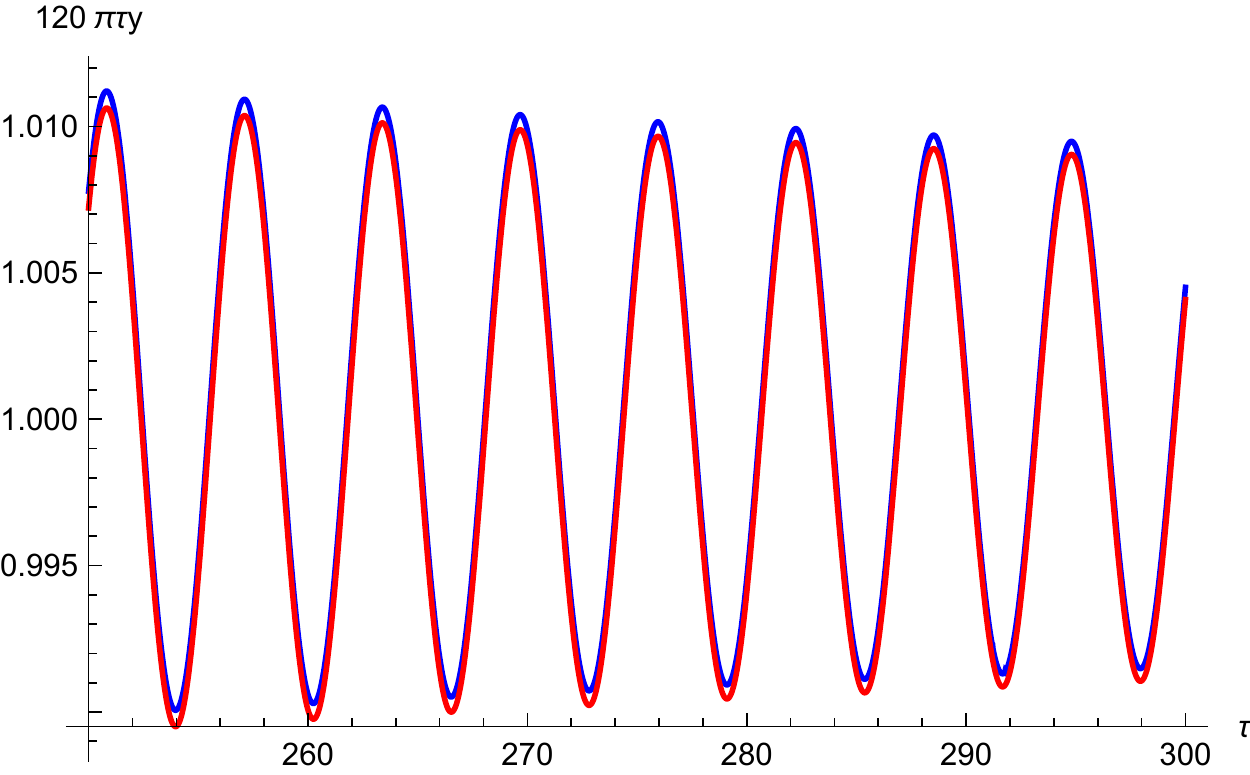}
    
  \end{minipage}
  \caption{
  Comparison of the numerical solution for the dimensionless energy density $120 \pi \tau \,y(\tau)$ (blue) for $w=1/3$
  with the asymptotic expression~(\ref{y-fin-asym}) (red) for moderately large $\tau$ (left panel)  and very large $\tau$ 
  (right panel). The agreement is very good.} 
\label{f:y-compar}
\end{figure}

Let us now solve numerically differential Eq.~(\ref{y-prime-w}) with $h$ and $r$ given by Eqs.~(\ref{hsol})
and (\ref{rsol}) and $\langle r^2 \rangle = 16/\tau^2$. The results are presented in Fig.~\ref{f:rho203}.
 \begin{figure}[!htbp]
  \centering
  \begin{minipage}[b]{0.3\textwidth}
    \includegraphics[width=\textwidth]{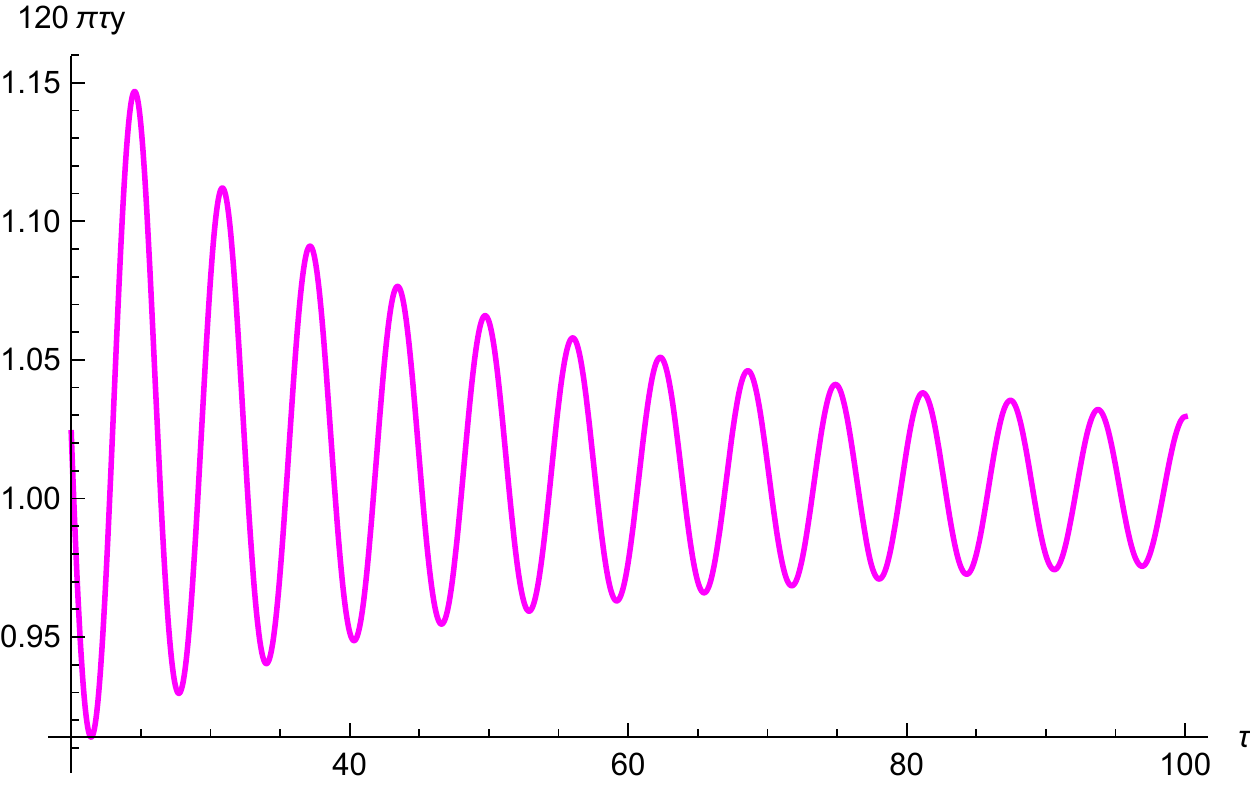}
      \end{minipage}
  \hfill
  \begin{minipage}[b]{0.3\textwidth}
    \includegraphics[width=\textwidth]{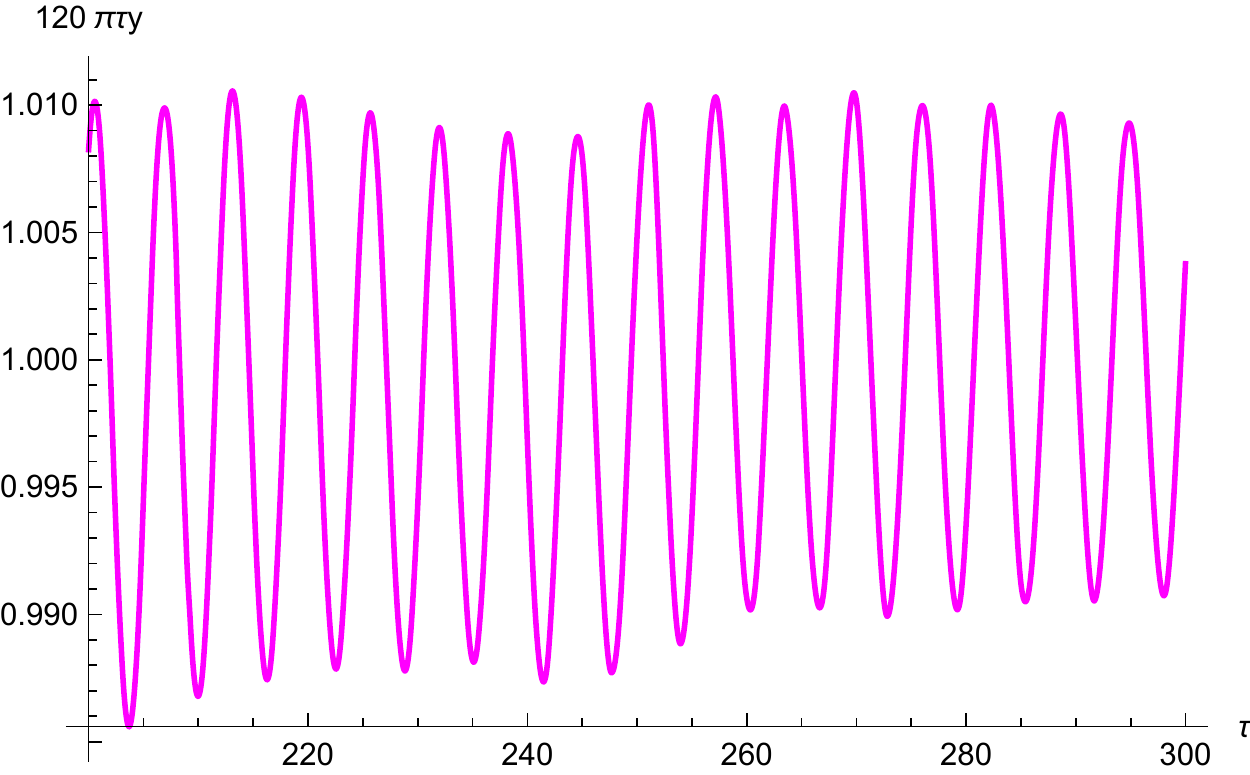}
\end{minipage}
 \hfill
  \begin{minipage}[b]{0.3\textwidth}
    \includegraphics[width=\textwidth]{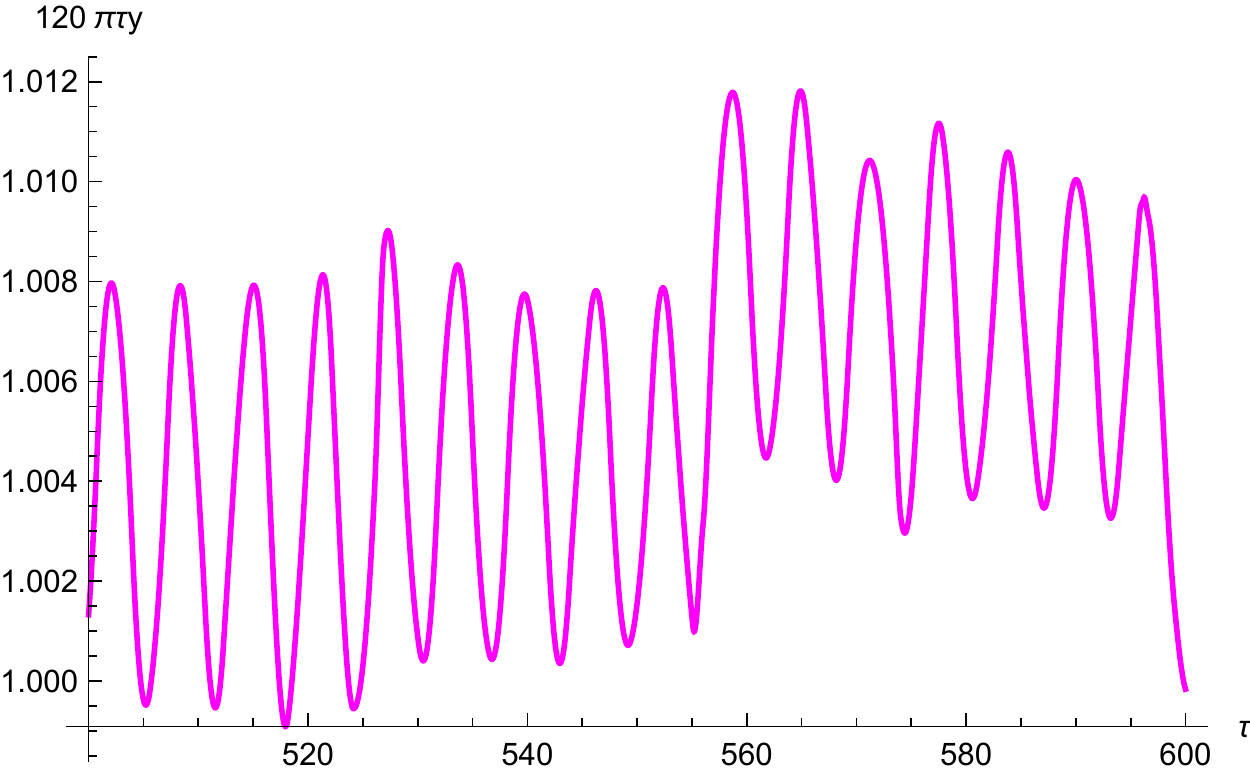}
\end{minipage}
 \caption{
 Numerical solution of Eq.~(\ref{y-prime-w}) for $120 \pi  \tau y(\tau) $ in different time intervals with $w =1/3$.
  }
  \label{f:rho203}
 \end{figure}
 Note that the numerical
 solution of the differential equation is not accurate at very large $\tau$. 
 On the other hand, the asymptotic expression does nor suffer from the mentioned shortcomings.

\subsection{Asymptotic solution at $\tau \gg 1$, $\gamma \tau \lesssim 1$,
 and $w=0$}\label{ss-asympt-0}

If $ w=0 $, equations (\ref{h-prime})-(\ref{y-prime}) take the form 
\be
h' + 2h^2 &=&  - r/6, \label{h-prime-2} \\
r'' + 3h r' + r &=& - 8 \pi \mu^2  y, \label{r-two-prime-2}\\
y' + 3h\,y &=& S[r], \label{y-prime-2}
\ee
Since $\mu \ll 1$ the impact of the r.h.s. in Eq.~(\ref{r-two-prime-2}) is not essential and we can use the 
expressions (\ref{hsol}) and (\ref{rsol}) for $h$ and $r$
from the previous subsection. Moreover the numerical solutions presented in Figs.~\ref{f:r-postinf} and \ref{f:hw}
strongly support this presumption. The only essential difference with the $w=1/3$ case arises in the 
equation (\ref{y-prime-2}) governing the evolution of the energy density. So to calculate $y(\tau)$ we can use slightly modified  results
of subsection~\ref{ss-asympt}. We need to solve equation  \eqref{y-prime-2}. Correspondingly there appears coefficient (-3) in the exponent, instead of (-4), as in Eq. (\ref{y-asympt-1/3}):
\be 
y(\tau) = \frac{1}{72\pi} \,\int_{\tau_0}^\tau \frac{d\tau_2}{\tau_2^2}\,\exp \left[ - 3 \int_{\tau_2}^\tau d\tau_1 h (\tau_1) \right].
\label{y-asympt-0}
\ee
Repeating the same calculations as in subsection~\ref{ss-asympt} we  obtain instead of Eq.~(\ref{y-fin-asym}):
\be 
y_{0} = \frac{1}{72\pi \tau} +\frac{ \cos \left( \tau + \theta\right)}{36 \pi \tau^2} .
\label{y-fin-asym0}
\ee 
It is instructive to compare the numerical calculations of integral~\eqref{y-asympt-0} with its asymptotic
 approximation \eqref{y-fin-asym0}. The results are presented in Fig.~\ref{f:rho-compar}. The agreement is impressive. 
 
 \begin{figure}[!htbp]
  \centering
  \begin{minipage}[b]{0.45\textwidth}
    \includegraphics[width=\textwidth]{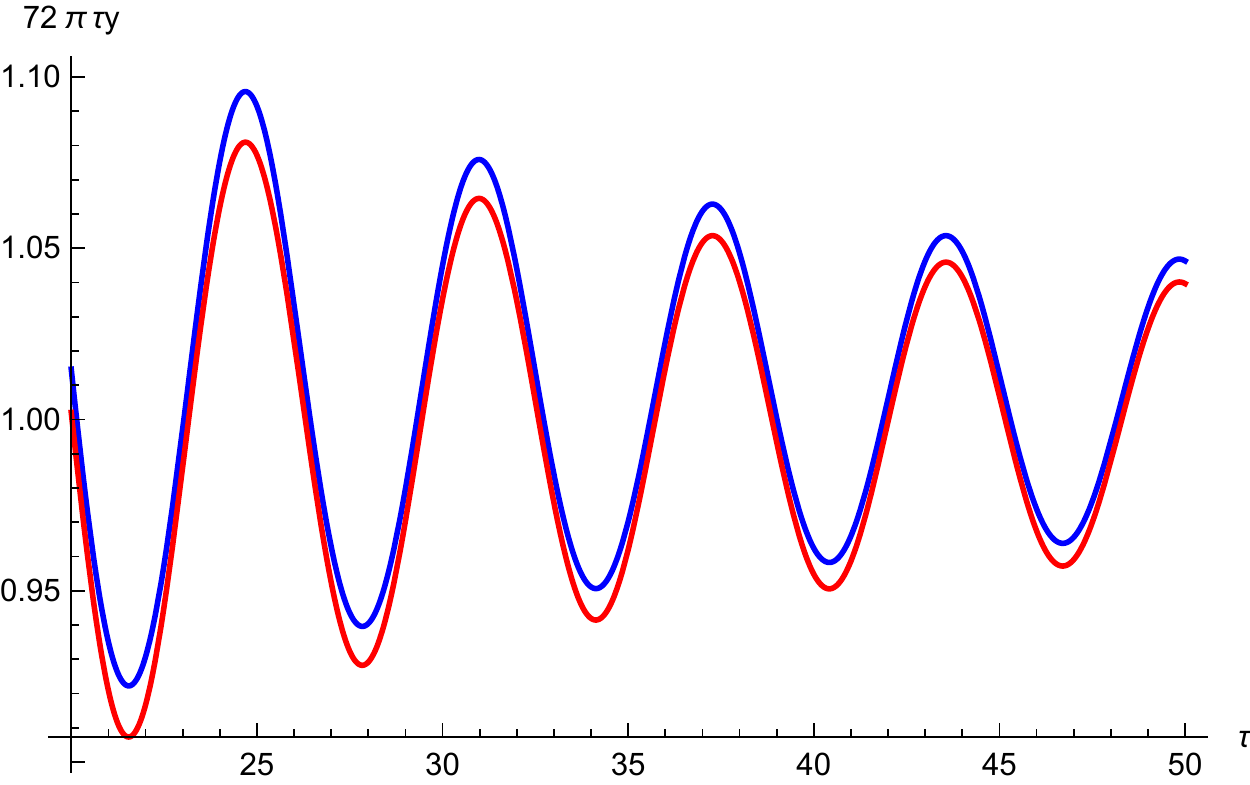}
    
  \end{minipage}
  \hfill
  \begin{minipage}[b]{0.45\textwidth}
    \includegraphics[width=\textwidth]{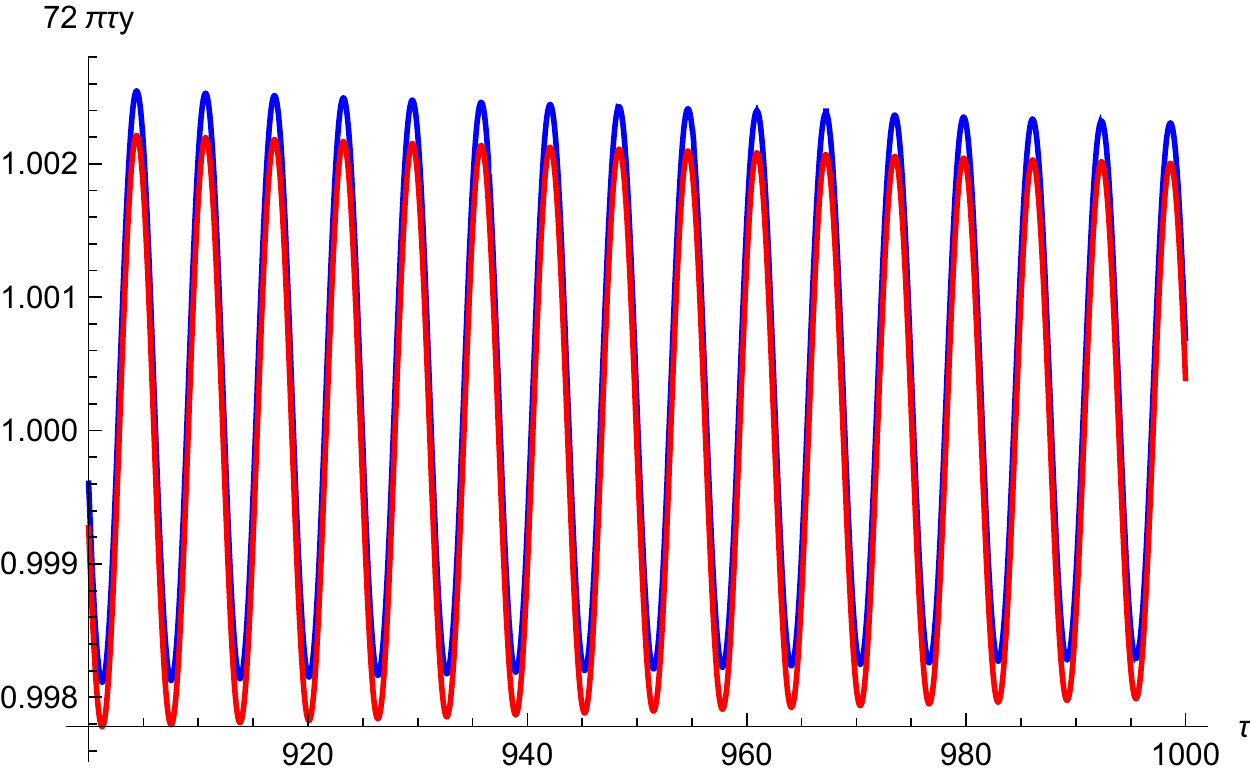}
    
  \end{minipage}
  \caption{
  Comparison of the solutions of the differential Eq.~\eqref{y-prime-2} for $w=0$: integral solution \eqref{y-asympt-0} (blue) and 
   asymptotic solution (\ref{y-fin-asym0}) (red). 
      The dimensionless energy density $72 \pi \tau \,y(\tau)$ is presented
   for moderately large $\tau$ (left panel)  and very large $\tau$ 
  (right panel). The agreement is very good.} 
\label{f:rho-compar}
\end{figure}

\subsection{Energy influx to cosmological plasma from the scalaron decay \label{ss-part-prod}}

Energy conservation demands that the energy influx to the cosmological plasma according to Eq.~\eqref{source} is 
ensured by the scalaron decay with the width $\Gamma$~\eqref{Gamma}. 
{ To check that let us consider a simplified model with the action of the scalar field $R$ of the form:
\be 
A_R = \frac{m_{Pl}^2}{48 \pi m^4}\,\int d^4x \sqrt{-g} 
\left[ \frac{(DR)^2}{2} - \frac{m^2 R^2}{2} - \frac{ 8\pi m^2}{m_{Pl}^2}\,  T^\mu_\mu R \right] .
\label{A-R}
\ee
which leads to the proper equation of motion~\eqref{D2-R}. To determined 
the energy density of the scalaron field 
we  have to redefine this field in such a way that the kinetic term of the new field enters the action 
with the coefficient unity. So the canonically normalized scalar field is:
\be
\Phi = \frac{m_{Pl}}{\sqrt{48 \pi}\, m^2}\,R.
\label{Phi}
\ee
Correspondingly, the energy density of the scalaron field is equal to: 
\be
\rho_R = \frac{\dot\Phi^2 +m^2 \Phi^2}{2} = \frac{m_{Pl}^2 (\dot R^2 + m^2 R^2)}{96 \pi m^4}.
\label{rho-R}
\ee
The energy production rate is given by:
\be 
\dot \rho_R = 2 \Gamma \rho_R = \frac{\dot R^2+ m^2 R^2}{2304 \pi  m} = \frac{m^3}{72 \pi t^2}
\label{dot-rho-R}
\ee
The coefficient 2 in front of $\Gamma$ appears because a pair of particles is produced in the 
scalaron decay. We take $\Gamma$ from Eq.~\eqref{Gamma}, for $R$ we use expression \eqref{rsol} and differentiate only
the quickly oscillating factor.


Let us compare  result~\eqref{dot-rho-R} with Eq.\eqref{dot-rho-pp} or ~\eqref{source}, 
if we take the amplitude of harmonic oscillations of $R$ equal to $R_{ampl} = 4m/t$ according to  Eq.~\eqref{rsol}. Correspondingly, the contribution of the particle production into Eq.~\eqref{dot-rho-pp} 
is exactly the same as above, as is to be expected:}
\be
\dot \rho_{source} =  \frac{m R^2_{ampl}}{1152\pi} = \frac{m^3}{72 \pi t^2}.
 \label{source-sol}
 \ee 

{ Note that our result for the width $\Gamma$ differs from those of Ref.~\cite{Gorb-Panin} by factor $1/\pi$. Without this factor the  
source~(\ref{source}) and $\Gamma$~\eqref{Gamma}  would be incompatible.}

One more comment is in order here.  Above, solving equations for the evolution of $R$ and $\rho$ we neglect particle
production effect for $R$, while took it into account in the equation for $\dot \rho$, despite both these effects having similar
magnitude. The reason for this approximation is that the energy density of the scalaron oscillations, $\rho_R$~\eqref{rho-R}, is 
large and the decrease of $\rho_R$ due to particle production is indeed relatively unimportant, while the cosmological 
plasma is completely created by the source term (\ref{source-sol}).

\subsection{Comments on the cosmological evolution at 
$\tau \lesssim 1/\gamma$ \label{ss-cosm-evol-small-tau} }

According to the results obtained above the cosmological evolution in $R^2$-gravity is strongly 
different the usual FRW-cosmology. Firstly, the energy density of matter in $R^2$ modified gravity
at RD stage drops down 
as (see Eq.~(\ref{y-fin-asym}) ):
\be
\rho_{R^2} = \frac{m^3}{120 \pi t} 
\label{rho-R2}
\ee
instead of the classical GR behavior 
\be
\rho_{GR} = \frac{3H^2m_{Pl}^2}{8 \pi} = \frac{3 m_{Pl}^2}{ 32 \pi t^2} .
 \label{rho-GR}
\ee
Secondly, the Hubble parameter quickly oscillates with time (\ref{hsol}), almost touching zero, and it is practically the same for RD 
and MD stages. And last but not the least, the curvature scalar drops down as $m/t$ and
oscillates changing sign (\ref{rsol}) instead of being proportional to the
trace of the energy-momentum tensor of matter, being identically zero at RD stage and monotonically decreasing with time, as 
$1/t^2$ at MD stage. It is noteworthy that $R$ is not related to the energy density of matter as is true in GR.

Because of this difference between the cosmological evolution in the
$R^2$-theory and GR the conditions for thermal 
equilibrium in the primeval plasma also very much differ. Assuming that the equilibrium with 
temperature $T$ is established, we estimate the particle reaction rate as
\be
\Gamma_{part} \sim \alpha^2 T,
\label{Gamma-par}
\ee
where $\alpha$ is the coupling constant of the particle interactions, typically $\alpha \sim 10^{-2}$. 
Equilibrium is enforced if $\Gamma_{part} > H$ or $\alpha^2 T t > 1$. The energy density of relativistic
matter in thermal equilibrium is expressed through the temperature as: 
\be
\rho_{therm} = \frac{\pi^2 g_*}{30}\, T^4 
\label{rho-therm}
\ee 
where $g_*$ is the number of relativistic species in the plasma. We take $g_* \sim 100$.

Using the equations \eqref{rho-R2} and \eqref{rho-therm}, we find that the equilibrium condition for $R^2$ cosmology is:
\be 
\left( \alpha^2  t T\right)_{R^2} =\frac{30 \alpha^2}{120\pi^3 g_*} \left(\frac{m}{T}\right)^3 =  8\cdot 10^{-9}  \left(\frac{m}{T}\right)^3 > 1. 
\label{Tt-R2}
\ee  
Analogously from \eqref{rho-GR} and \eqref{rho-therm} it follows that for GR-cosmology:
\be 
\left(\alpha^2  t T\right)_{GR} =\alpha^2 \left(\frac{90}{32\pi^3 g_*}\right)^{1/2} \frac{m_{Pl}}{T}
=  3\cdot 10^{-6} \,\frac{m_{Pl}}{T} > 1. 
\label{Tt-GR}
\ee  
Correspondingly equilibrium between light particles in $R^2$-cosmology is established, when $T_{R^2} < 2\cdot 10^{-3} m$,
while in GR the corresponding condition is $ T_{GR} < 3\cdot 10^{-6} m_{Pl}$.

{ Let us stress that  expressions~(\ref{Tt-R2}) and in (\ref{Tt-GR}) determine the temperature below which thermal
equilibrium is established in the primeval plasma. This temperature is not the same as the so called heating temperature $T_h$. 
The latter is defined by the condition that all energy of the scalaron field is transferred into the energy of the plasma. This takes
place approximately at $t\Gamma = 1$. Correspondngly
\be
T_h \approx \frac{m}{(192\pi^2)^{1/4}}\,\sqrt{\frac{m}{m_{Pl} }}
\label{T-h}
\ee
For $m = 3 \times 10^{13}$ Gev $T_h \approx 6\times 10^8$ GeV, which is close to other estimates presented in the literature.
If we take into account a possible delay of the scalaron decay by $\ln (m/\Gamma)$-factor, see the very end of 
Sec.~\ref{s-large-gamma-tau}, $T_h$ would be sightly lower.
}

Let us note that during inflation the curvature $R(t)$ did not oscillate. The particle production started after the onset of the
oscillations at $t \gg 1/m$. So the energy density of the cosmological plasma never exceeded $m^4/(120 \pi)$ and correspondingly 
its temperature is bounded by $T_{R2} \lesssim 0.01 m$. Moreover,  the energy of the individual particles created by the scalaron 
oscillations, as well as their masses, must be smaller than $m/2$.  This fact, in particular, opens a possibility to make dark matter (DM)
particles from the lightest supersymmetric particles (LSPs). According to the LHC data the mass of LSP must be above 
TeVs. In this case the cosmological energy of LSP would be much higher than the observed density of DM, if LSPs are
thermally produced. In $R^2$ cosmology the production of heavy LSPs could be strong enough suppressed for $m_{LSP} > m$,
so the energy density of LSP may be sufficiently low.

$R^2$ cosmology may also noticeably change the probability of primordial black hole formation and predictions for high temperature 
baryogenesis, in particular, baryo-through-lepto genesis. These problems are outside the frameworks of the present work and will be
considered elsewhere.

\section{Solution at $\gamma \tau \gtrsim 1$  \label{s-large-gamma-tau}}
 
We consider the system of equations (\ref{h-prime})-(\ref{y-prime}). Unfortunately a
straightforward numerical solution of this system of equations quickly becomes unreliable  because  the standard 
Mathematica program does not properly evaluate very small exponential suppression factor 
$\exp{(-\gamma \tau/2)}$, when $\gamma\tau \gg 1$.
So we need to proceed differently.
Firstly, as in the previous Section, we study the case of relativistic matter, i.e. $w = 1/3$.  
Eliminating the first derivative $r'$ from Eq.~(\ref{r-two-prime}) by introducing the new function
$v$ according to:
\be 
r = \exp \left[-\gamma (\tau-\tau_0)/2 - (3/2)\,\int_{\tau_0}^\tau d\tau_1  h(\tau_1) \right] v  ,
\label{r-of-v}
\ee
we come to the equation
\be
v'' + \left[ 1 - \frac{(\gamma + 3h)^2}{4} \right] \, v = 0 .
\label{v-two-prime}
\ee
Since in realistic case $\gamma \ll 1$ and $h \lesssim \gamma$, because  $h \sim 1/\tau$ 
and by assumption $ \gamma \tau \gtrsim 1$, the second term in the square brackets can be neglected and
Eq.~(\ref{v-two-prime}) is trivially solved as:
\be
v = v_0 \,\cos \left( \tau - \tau_0 + \theta_v \right),
\label{v-sol}
\ee
where the amplitude $v_0$ can be approximately determined by matching of the solution \eqref{r-of-v} to Eq.~(\ref{rsol})
at $\gamma \tau_0 \sim 1$, considered in the previous section. Thus
\be
v = -4\gamma \,\cos \left( \tau + \theta \right),
\label{v-sol2}
\ee
According to  Eq.~(\ref{r-of-v}) the curvature $r$  exponentially vanishes at large $\gamma \tau /2$, so the
r.h.s. of Eq.~(\ref{h-prime}) tends to zero with the same speed, and the Hubble parameter approaches
$1/(2\tau)$, as it, indeed, takes place in the standard cosmology at RD stage. The energy density in this limit satisfies
Eq.~(\ref{y-prime}) with vanishing  r.h.s., thus $y$ drops down as $1/a^4$ as expected. However, it is not
clear from these equations if the standard relation between $H$ and $\rho$:
\be
H^2 = \frac{8\pi}{3} \, \frac{\rho}{m_{Pl}^2} 
\label{H-of-rho}
\ee
is fulfilled.
To see that, we need the Friedmann-like equation for the 00-component of the $R^2$-modified gravity equation~\eqref{field_eqs}
in the limit when particle production can be neglected. This equation can be written as:
\be 
H^2 +\frac{1}{m^2} \, \left[ 2 \ddot H H - \dot H^2 + 6\dot H H^2  \right] 
= \frac{8\pi \rho}{3 m_{Pl}^2} .
\label{h-test-1}
\ee
According to our estimate the curvature exponentially disappeared at $\gamma \tau > 1$, and, correspondingly,  
$ \dot H + 2 H^2 =0 $.
In this case the term in square brackets of Eq.~\eqref{h-test-1} vanishes and  the normal cosmology is restored.

More interesting is the case of nonrelativistic dominance, $w = 0$, or some deviations from the strict $w=1/3$
due to presence of massive particles in the cosmological plasma or due to conformal anomaly.
Now we have to study Eq.~(\ref{r-two-prime}) with non-zero r.h.s. which might change the asymptotical 
exponential decrease of $r$. Making the transformation (\ref{r-of-v}) and neglecting the minor term $(\gamma + 3h)^2/4$
in Eq.~\eqref{v-two-prime}, we arrive to
\be
v'' + v =  - 8 \pi \mu^2 (1-3w) y(\tau)\, \exp \left[\gamma (\tau-\tau_0)/2 + (3/2)\,\int_{\tau_0}^\tau d\tau_1  h(\tau_1) \right] .
\label{v-rhs}
\ee
The value of $(1-3w)$ is not yet specified here, we only assume that it is nonzero.

This equation is solved as:
\be
v(\tau)= -8\pi \mu^2(1-3w)\int_{\tau_0}^\tau d\tau_1 \sin(\tau-\tau_1) y (\tau_1) 
\exp  \left[\frac{\gamma (\tau_1-\tau_0)}{2} + \frac{3}{2}\int_{\tau_0}^{\tau_1} d\tau_1  h(\tau_1) \right] ,
\label{v-anal}
\ee
plus a solution ~(\ref{v-sol}) of the homogeneous equation (\ref{v-two-prime}), 
as it can be easily checked by direct substitution.

Now using relation (\ref{r-of-v}) we find for the curvature scalar:
\be
r = -8 \pi \mu^2 (1-3w) \int_{\tau_0}^{\tau} d\tau_1 y(\tau_1)  \sin(\tau - \tau_1)\exp{\left[-\frac{\gamma}{2} (\tau-\tau_1) 
-\frac{3}{2} \int_{\tau_1}^\tau d\tau_2 h (\tau_2)  \right] }+r_{h} ,
\label{r-test-GR}
\ee
where $r_{h}$ is a  solution of the homogeneous equation:
\be
r_{h} =  r_0 \cos(\tau + \theta_r)\,\exp\left[-\frac{\gamma}{2} (\tau-\tau_0) 
-\frac{3}{2} \int_{\tau_0}^\tau d\tau_2 h (\tau_2)  \right] .
 \label{r-h}
\ee
It is noteworthy that the solutions of the homogeneous
equation drop down exponentially as $\exp (-\gamma \tau/)2$, while the inhomogeneous part does not, since integral \eqref{r-test-GR} is dominated
by $\tau_1$ close to $\tau $.

Let us assume that the standard General Relativity became (approximately) valid after sufficiently long 
cosmological time and check if this assertion
is compatible with Eq.~(\ref{r-test-GR}). So we take, according to the standard cosmological laws with $w >-1$:
\be 
a \sim t ^{\frac{2}{3(1+w)}},\,\,\,\, H = \frac{2}{3(1+w) t}, \,\,\,\, 
\rho = \frac{3H^2 m_{Pl}^2}{8\pi}= \frac{m_{Pl}^2}{6\pi (1+w)^2 t^2}  .
\label{st-cosm}
\ee

Introducing new integration variables $ x = \tau_1 /\tau$, $ x_2=\tau_2/\tau$ and taking integral 
over $dx_2$ we obtain:
\be 
r_{inh} = -\frac{4 (1-3w)}{3 (1+w)^2 \tau} \int_\epsilon^1 dx \,
{\sin \left[ \tau(1-x) \right]}\, \left(\frac{1}{x}\right)^{\frac{1+2w}{1+w}}\,
\exp\left[ - \frac{\gamma \tau}{2} (1-x) \right] ,
\label{r-test-GR2}
\ee
where  $r_{inh}$ is the contribution to $r$ from the inhomogeneous term in the equation of motion and $\epsilon = \tau_0 /\tau \ll1$. 
For large $\tau$ and $\gamma \tau$ the integral in Eq.~(\ref{r-test-GR2}) can be estimated in the same way as
the integral~(\ref{I}). First we write 
\be
\sin 
\label{sine} \left[ \tau(1-x) \right] = \frac{i}{2} \left[ e^{-i \tau (1-x)} - e^{i \tau (1-x)}\right].
\ee
The integral with the first exponent can be reduced to the difference of
two integrals along the contours $ x = \epsilon + i\zeta$  and 
$ x = 1 + i\zeta$, where $\zeta$ runs from 0 to infinity, while that with the second exponent 
is expressed through the similar integrals along the negative $\zeta$ axis. So we obtain:
\be
r_{inh}= \frac{2 (1-3w)}{3 (1+w)^2 \tau} \int_0^\infty d\zeta \, e^{-\tau \zeta} 
&&\left[  \left(\frac{1}{\epsilon +i\zeta}\right)^{\frac{1+2w}{1+w} }
\exp{ \left(-i \tau (1-\epsilon) - \frac{\gamma \tau (1 -\epsilon - i \zeta)}{2}\right) } \right. \nonumber \\
&&- \left. \left(\frac{1}{1 +i\zeta}\right)^{\frac{1+2w}{1+w} } \exp \left( i\frac{\gamma \tau  \zeta}{2}\right) 
 + h.c.\right] .
\label{r-asym-1}
\ee
Here "h.c" means "hermitian conjugate".

Since $\tau \gg 1$, the integrals effectively "sit" at $\zeta \sim 1/\tau$ because of $\exp (-\zeta \tau)$ factor.
In the leading order the second term in the 
square brackets is equal to $(-1)$ and together with the hermitian conjugate after integration they give 
 $-2/\tau$. The first term is exponentially suppressed at large $\gamma\tau/2$, 
as $\sim \exp (-\gamma\tau/2)$.  Note that there is a large pre-exponential factor proportional to
$\tau^{(1+2w)/(1+w)}$,
which slows down the approach to the  GR value, presumably reached asymptotically:
\be
R = - \frac{8\pi }{\mpl^2}(1 - 3w)\rho  = \frac{4(1-3w)}{3(1+w)^2 t^2} \,.
\label{R-GR}
\ee

However, our input functions do not coincide with those found at large 
$\tau$ but small $\gamma \tau\lesssim 1$,
given by Eq.~(\ref{hsol}). We make a more general anzatz,
taking the dimensionless Hubble parameter and energy density as
\be
h_{test} (\tau) = \frac{h_1 + h_2 \sin (\tau +\theta_h) }{\tau},\,\,\,\ y(\tau) =\frac{y_1}{\tau^\beta}
\label{h-test}
\ee
and check if it is possible to adjust the constants $h_1$, $h_2$, $y_1$, and $\beta$ to restore GR. 
Correspondingly from Eq.~(\ref{r-test-GR}) we obtain
\be
 r_{inh} = -\frac{8 \pi \mu^2 y_1(1-3w) }{ \tau^{\beta -1}}  \int_\epsilon^1 dx_1 \,
{\sin \left[ \tau(1-x_1) \right]}\, \left(\frac{1}{x_1}\right)^{\beta-3h_1/2} \times  \nonumber \\
\exp \left[ \left( - \frac{\gamma \tau}{2} (1-x_1) \right) -\frac{3h_2}{2} \int_{x_1}^1 \frac{dx_2}{x_2}\sin(\tau x_2 +\theta_h)
\right] ,
\label{r-of-h1-h2}
\ee
Let us first estimate  the integral over $dx_2$:
\be 
I_2 \equiv \int_{x_1}^1 \frac{dx_2}{x_2}\sin(\tau x_2 +\theta_h) 
\label{I2-x1}
\ee
The result depends upon the lower limit of the integration, $x_2 = x_1$.  Let us estimate $I_2$ in the same way as it
is done with Eq.~(\ref{r-asym-1}), i.e. integrate along  the two contours 
$x_2 = 1 \pm i\zeta_2$ and  $x_2 = x_1 \pm  i\zeta_2$. After straightforward calculations we obtain
\be
I_2 = - \frac{\cos (\tau +\theta_h )}{\tau} + 
\int_0^\infty d\zeta_2\, e^{-\tau\zeta_2} \left[ 
\frac{x_1 \,\cos (x_1 \tau + \theta_h)}{x_1^2+\zeta_2^2}  +
\frac{\zeta_2 \,\sin (x_1 \tau + \theta_h)}{x_1^2+\zeta_2^2} 
\right] .
\label{I2-small-x1}
\ee
The first term comes from the contour $x_2 = 1 \pm i\zeta_2$, while the integral over $d\zeta_2$
comes from $x_2 = x_1 \pm  i\zeta_2$.
Here the effective value of $\zeta_2$ is small, $\zeta_2 \sim 1/\tau$, due to the factor $\exp (-\zeta_2 \tau)$.

The integral over $dx_1$  in Eq.~(\ref{r-of-h1-h2}) runs in the limits $\epsilon < x_1 < 1$. Hence in Eq~(\ref{I2-small-x1})
we may assume that $x_1 \gg \zeta_2$. Indeed the effective value of $\zeta_2 $ is about $1/\tau$, while 
$\epsilon = \tau_0 /\tau$ with $\tau_0 \gg 1$. So we find:
\be
I_2 \approx - \frac{\cos (\tau +\theta_h )}{\tau} +  \frac{\cos ( x_1 \tau +\theta_h )}{x_1\tau} + \frac{const}{\tau^2}
\label{I-2-fin}
\ee
and can conclude that $I_2 \ll 1$ and $\exp (-3h_1 I_2 /2) \approx 1$. Thus:
\be
 r_{inh} \approx -\frac{8 \pi \mu^2 y_1(1-3w) }{ \tau^{\beta -1}}  \int_\epsilon^1 dx_1 \,
{\sin \left[ \tau(1-x_1) \right]}\, \left(\frac{1}{x_1}\right)^{\beta-3h_1/2}   
\exp  \left[ - \frac{\gamma \tau}{2} (1-x_1) \right]
\label{r-fin}
\ee

The integration over $dx_1$ in Eq.~ (\ref{r-fin}) can be transformed, as above, to the integrals along the two contours
$x_1 = \epsilon \pm i\zeta$ and  $x_1 = 1 \pm i\zeta$. The result is similar to Eq.~(\ref{r-asym-1}) but the parameters
$\beta$ and $h_1$ are not expressed through $w$ but at this stage remain free:
\be 
r_{inh}  \approx \frac{8 \pi \mu^2 y_1(1-3w) }{ \tau^{\beta -1}} 
\int_0^\infty d\zeta \, e^{-\tau \zeta} 
&&\left[  \left(\frac{1}{\epsilon +i\zeta}\right)^{\beta -\frac{3h_1}{2} }
\exp{ \left(-i \tau (1-\epsilon) - \frac{\gamma \tau (1 -\epsilon - i \zeta)}{2}\right) } \right. \nonumber \\
&&- \left. \left(\frac{1}{1 +i\zeta}\right)^{\beta -\frac{3h_1}{2} } \exp \left( i\frac{\gamma \tau  \zeta}{2}\right) 
 + h.c.\right] .
\label{r-22}
\ee
Keeping in mind that $\zeta \sim 1/\tau$ and that $\epsilon= \tau_0 /\tau \gg 1/\tau$ we simplify the result as:
\be
r_{inh} \approx  \frac{16 \pi \mu^2 y_1(1-3w) }{ \tau^{\beta }}
\left[ \left( \frac{\tau}{\tau_0}\right)^{\beta -3h_1/2} e^{-\gamma (\tau-\tau_0)/2}  \cos (\tau -\tau_0) -1
\right] .
\label{r-fin-23}
\ee

This expression is a small correction to homogeneous solution (\ref{rsol}), for which $\beta =1$, $y_1 = 1/(72\pi)$ from Eq.~\eqref{y-fin-asym0},
 and $h_1 = 2/3$.
The first term in square brackets at large $\tau $, but small $\gamma \tau $,
has the same the same dependence on $\tau$ as 
 Eq.~(\ref{rsol}), but here the coefficient depends upon $w$. 
Ultimately, at $\gamma \tau > 1$, the first term dies down and only the last non-oscillating term survives. In this
limit the particle production by $R$ vanishes, or strongly drops down. 

According to Eq. (\ref{y-prime}) in the absence of particle production the dimensionless energy density 
drops down as $y \sim 1/a^{3(1+w)}$. Since the oscillations exponentially disappear the derivatives of $H$ in 
Eq.~(\ref{h-test-1}) can be neglected and $H$ satisfies the GR relation~(\ref{H-of-rho}) with $\rho =m^4 y$ decreasing as $1/\tau^2$
independently of the value of $w$.

The transition from the modified $R^2$-regime to GR is due to the inhomogeneous part of the solution for $r$ which 
does not drop down exponentially, i.e. due to the last term in the square brackets of Eq.~(\ref{r-fin-23}). It is natural to expect 
that the GR regime starts roughly  at $\tau \gtrsim 1/\gamma$. We may make a simple estimate using Eq.~(\ref{r-fin-23})
with $w =0$. In this case we have to compare the value of the curvature scalar $r =  2\mu^2/(9\tau)$ with
homogeneous solution for the curvature (\ref{r-h}): $r \sim 4\exp (-\gamma\tau/2) /\tau$. These two expressions 
become comparable at  $\gamma \tau \approx 2 \ln(1/\mu^2) $, where $  \ln(1/\mu^2) $ may
 be much larger than unity. Unfortunately 
similar arguments cannot be applied to $w=1/3$ because the GR curvature in this case is identically zero. 
In realistic  case $w$ differs from zero either due to presence of massive particles in the primeval plasma or
because of the conformal anomaly. Such estimation of the moment of transition to GR looks very unnatural. 
Probably an analysis of all equations of motion may permit  the same result for the transition to GR
time $\gamma \tau \sim \ln (1/\gamma)$.


\section{Conclusion \label{s-conclud}}

Cosmological history in $R^2$-modified gravity can be separated into four distinct epoch. 
It started from an exponential (inflationary) expansion. At this stage the universe was void and dark with slowly 
decreasing curvature scalar $R(t)$. The initial value of $R$ should be quite large, $R >  300 m^2$ to ensure sufficiently 
long inflations, such that the number of e-folding exceeded 70.

Next epoch began when $R$ dropped down to zero and started to oscillate around it
as $ R \sim m \cos (m t) / t$. 
The curvature oscillations resulted in the onset of particle production and this moment can be called Big Bang.
The universe expansion at this stage is described by very simple, but unusual law with the Hubble parameter periodically 
reaching (almost) zero,  $H = (2/3t) [1 + \sin (mt)]$ (\ref{hsol}). Such a regime was realised asymptotically for large time, $m t \gg1$,
but $\Gamma t \lesssim 1$.

Later, when time becomes so large that $\Gamma t$ exceeds unity, the oscillations of all relevant quantities exponentially damps down 
and the particle production by curvature switches off, becoming negligible. This is the transition period to General Relativity. Presumably 
it takes place when $\Gamma t $ becomes larger than unity by the logarithmic factor, $\ln (m_{Pl}/m)$.  

After this time we arrive to the usual GR cosmology. Requesting that the GR regime started before Big Bang Nucleosynthesis, 
we find the limit $m > 10^5 $ GeV~\cite{ADR-R2}. 
{ According to Ref.~\cite{Faulkner:2006ub}, comparison of theoretical prediction for
density perturbations generated during $R^2$-inflation~\cite{Starobinsky_1980} with the CMB fluctuation data and the large scale 
structure leads to the conclusion that $m \approx 2\times 10^{12}$ GeV. The analysis made in the subsequent 
works~\cite{Gorb-Panin} leads to the somewhat different  result  $m \approx 3\times 10^{13}$ GeV. More complicated models
of inflation based on different GR modification could lead to considerably different value of the normalization mass $m$.}
}

Unusual cosmological evolution during the time $t < 1/\Gamma$ would lead to noticeable modification of the cosmological baryogenesis
scenarios, to a variation of the probability of formation of primordial black holes, and 
to the a change of the frozen density of dark matter particles. In particular, it opens window for heavy lightest supersymmetric 
particles to be the cosmological dark matter.

\section*{Acknowledgment}

EA and AD acknowledge the support of  the RSF Grant N 16-12-10037.

  \end{document}